\documentclass[aps,amssymb,reprint,twocolumn,superscriptaddressgroupaddress]{revtex4-1}
\usepackage{amsmath}
\usepackage{hyperref}
\usepackage[english]{babel}
\usepackage[utf8]{inputenc}
\usepackage{float}
\usepackage{graphicx}
\usepackage{braket}
\usepackage{xcolor}
\allowdisplaybreaks

\begin{document}
\title{One-dimensional few-electron effective Wigner crystal \\in quantum and classical regimes}
\author{DinhDuy Vu}
\author{S. Das Sarma}
\affiliation{Condensed Matter Theory Center and Joint Quantum Institute, Department of Physics, University of Maryland, College Park, Maryland 20742, USA}

\begin{abstract}
	A system of confined charged electrons interacting via the long-range Coulomb force can form a Wigner crystal due to their mutual repulsion. This happens when the potential energy of the system dominates over its kinetic energy, i.e., at low temperatures for a classical system and at low densities for a quantum one. At $T=0$, the system is governed by quantum mechanics, and hence, the spatial density peaks associated with crystalline charge localization are sharpened for a lower average density. Conversely, in the classical limit of high temperatures, the crystalline spatial density peaks are suppressed (recovered) at a lower (higher) average density. In this paper, we study those two limits separately using an exact diagonalization of small one-dimensional (1D) systems containing few ($<10$) electrons and propose an approximate method to connect them into a unified effective phase diagram for Wigner few-electron crystallization. The result is a qualitative quantum-classical crossover phase diagram of an effective 1D Wigner crystal. We show that although such a 1D system is at best an effective crystal with no true long-range order (and thus no real phase transition), the spatial density peaks associated with the quasi-crystallization should be experimentally observable in a few-electron 1D system. We find that the effective crystalline structure slowly disappears with both the crossover average density and crossover temperature for crystallization decreasing with increasing particle number, consistent with the absence of any true long-range 1D order. Thus, an effective few-electron 1D Wigner crystal may be construed either as existing at all densities (manifesting short-range order) or as non-existing at all densities (not manifesting any long-range order). Within one unified description, we show through exact theoretical calculations how a small 1D system interacting through the long-range Coulomb interaction could manifest effective Wigner solid behavior both in classical and quantum regimes. In fact, one peculiar aspect of the effective finite size nature of 1D Wigner crystallization we find is that even a short-range interaction would lead to a finite-size 1D crystal, except that the crystalline order vanishes much faster with increasing system size in the short-range interacting system compared with the long-range interacting one.
\end{abstract}
\maketitle
\section{Introduction}
The goals of the current work are to theoretically calculate the spatial density structure of a small collection of one-dimensional (1D) electrons interacting via the long-range Coulomb force and to determine how this structure depends on the electron average density ($\rho$), the temperature ($T$) as well as on the number of electrons ($N$) in the system. We also include a parameter in the Coulomb interaction, which mimics the short-distance cut-off caused by the transverse dimension of the physical system (e.g. the diameter of a carbon nanotube or a semiconducting quantum wire). Using exact diagonalization and statistical mechanics, we obtain results for both $T=0$ (quantum) and high-$T$ (temperature much higher than the Fermi temperature) classical situations, and then propose a smooth interpolation between quantum and classical regimes, through which we construct an effective crystallization phase diagram which should be valid at any temperatures and average densities. An important aspect of our results is a subtle electron number dependence of 1D Wigner crystallization, which arises from the Luttinger liquid nature of 1D systems. The subject of  effective 1D Wigner crystallization is well-established, going back to the 1980s.  Although many of the results we present here are known in the literature in different contexts and using different approximations, our having all of them together in one place, using exact theoretical techniques, covering both classical and quantum regimes as well as spinful and spinless systems and long- and short-range interactions for comparison, should serve a useful purpose.  

One-dimensional electron systems are special because the non-interacting Fermi surface is just two discrete points at $\pm k_F$. In this case, the bosonization method proves useful in solving exactly the corresponding interacting problem, leading to the concept of a Luttinger liquid \cite{Luttinger}. A Luttinger liquid is a paradigm for a non-Fermi-liquid as its momentum distribution function for the interacting 1D system is smooth and continuous through $k_F$ even at $T=0$ instead of having a finite discontinuity which is the hallmark of 2D and 3D Fermi liquids. At a finite temperature, it is not so easy to distinguish a Luttinger liquid from a finite temperature Fermi liquid as a practical matter since both have smooth momentum distribution functions through $k=k_F$ although, as a matter of principle, the two are very different \cite{finiteT1,finiteT2}. 

In the original work and most of the subsequent works on Luttinger liquids, the electron-electron interaction is assumed to be short-ranged since the singular nature of 1D interacting systems is essentially independent of the range of their mutual interaction \cite{shrange,shrange2,shrange3,shrange4}. Including long-range inter-electron Coulomb interactions in the interacting 1D spinful model, Schulz showed in a seminal work in 1990 that there exists a $4k_F$ oscillation in the spatial density correlations whose spatial decay rate is much slower than any power laws \cite{Schulz}. This makes the existence of a length-dependent effective Wigner crystal possible in 1D because the $4k_F$ period corresponds to an effective crystalline structure since $4k_F=2\pi/a$ in a spinful 1D electron system where $a=1/\rho$ is the average inter-particle separation. The 1990 theoretical work by Schulz could be considered the starting point of the subject matter of 1D Wigner crystallization which is the topic of the current work. Such a 1D Wigner crystal is obviously only a quasi-crystal since the oscillation dies out eventually, but there should be observable consequences of the slowly-decaying $4k_F$ oscillations in finite 1D Coulomb systems. The interplay between the apparent existence of a 1D Wigner crystal for finite systems and the eventual disappearance at long distances is an important theme of the current work.
 
 In practice, the 1D Luttinger liquid has been studied experimentally in effective 1D systems such as quantum wires \cite{quantumwire}, carbon nanotubes \cite{nanotube1} and organic conductors \cite{organic}. However, observing an effective 1D quasi-Wigner crystal is highly challenging since it must necessarily involve a small system (because of the absence of any true long-range order) and one must ensure that the environment containing the 1D electron system is free of disorder and that the density probing process is non-invasive on the system. These pristine experimental conditions for long-range interaction in a 1D system were recently met in Ref. \cite{exp0}, where a few electrons ($<10$) were confined in a clean carbon nanotube and a second nanotube was used as a non-invasive scanning probe to measure the spatial charge distribution of the electrons with minimal perturbation. The resultant 1D electrons spatial charge density at very low temperatures exhibits features of a quantum quasi-Wigner crystal, i.e., spatially equidistant  density peaks instead of a uniform liquid-like density distribution. Our work is motivated by this low-temperature nanotube experiment imaging the 1D quasi-Wigner crystal formation.  Unlike higher (2D or 3D) dimensional $T=0$ systems, the 1D Wigner crystal formation does not have a critical density associated with it.  Thermal fluctuations, however, destroy these local density correlations, and eventually the system should cross over to the corresponding classical Wigner crystal for $T>T_F$. Such a classical electron Wigner crystal (and the corresponding 2D liquid to solid classical transition with decreasing temperature) was observed in a system of 2D electrons confined on the surface of liquid helium a long time ago \cite{wigner2d}. Our goal in the current work is to do both $T=0$ and finite $T$ calculations to connect the 1D effective Wigner crystallization between  quantum and classical regimes.  We emphasize that, since there cannot be any true long-range 1D order, our results apply only to finite systems where a quasi-long-range order is meaningful.

 Although our work has been motivated by a recent experimental work \cite{exp0}, our goal is purely theoretical. We do not make any attempt to make quantitative contact with any experimental results, and indeed such a comparison between theory and experiment is unfeasible because of the effective nature of the 1D Wigner crystal - nothing sharp or decisive happens at any value of the system parameter, so a quantitative comparison is meaningless. Therefore, from now on, we consider an ideal 1D system with a generic interaction that resembles the asymptotic form $1/x$ of the real Coulomb interaction. We also consider the electron kinetic energy to be of the standard parabolic form. We are aware that there is a vast literature on the subject, investigating different aspects of quasi-1D systems using various simulation methods and microscopic calculations \cite{coulomb1, coulomb2, coulomb4, R2,R3,R4,R5,R6, spin_orbit}. The smallest possible Wigner crystals consisting just of two electrons, also called Wigner molecules, have also been studied in great details \cite{nanotube2, R1, wignermol}. Our work, on the other hand, attempts to exhibit many aspects of effective 1D Wigner crystallization within one unified abstract model, without getting into details of the experimental systems which should not be relevant for the fundamental theoretical picture. The simplicity of our generic theoretical model now allows us to examine the intertwined effects of the basic parameters of the 1D interacting model on the observation of an effective 1D Wigner crystal. In addition, we also study the thermal melting of the effective 1D Wigner crystal by smoothly interpolating between quantum and classical regimes. As a result, we show that there is an isolated phase of observable 1D Wigner crystal and this phase shrinks extremely slowly with increasing number of particles, which is consistent with the fact that is there is no true long-range order in 1D systems.  Again we note that rigorous simulations have been done to study the effect of temperature and produce the classical phase diagrams of Wigner crystal formation \cite{phase_montecarlo, finitetemp}. However, our simple phase diagram can be readily adapted and fine-tuned for a wide range of experimental parameters.  Our conclusions can then be translated qualitatively to real physical systems. Also, the quantum-classical crossover and the slow disappearance of the apparent 1D Wigner phase are other results of our work.
 
 In this paper, we investigate the spatial electron density profile in a 1D Coulomb system of $N$ ($=2-8$) electrons (with length scale $L$ and average density $\rho\approx N/L$) in the two limits of zero (quantum) and high (classical) temperatures. At $T=0$, the kinetic energy roughly scales as $L^{-2}$; and the Coulomb repulsive potential energy scales as $L^{-1}$. Consequently, with increasing $L$ or decreasing $\rho$, Coulomb repulsion becomes dominant, producing well-resolved spatial density peaks as the electrons attempt to stay away from each other. By contrast, in the high-temperature classical limit, the kinetic energy depends on the temperature and goes as $\sim k_BT$. Thus, on expanding the system size (or equivalently decreasing the average density) in this classical regime, the Coulomb repulsion decreases while the kinetic energy stays constant if $T$ is fixed. Therefore, the system becomes less crystalline at lower average densities in the classical regime in contrast to the quantum situation. Our goal is to connect these two opposite behaviors by using phonon vibration modes to estimate the ratio between the vibration amplitude and the average inter-electron spacing. Using an effective Lindemann melting criterion, we are able to produce a qualitative phase diagram of the effective 1D Wigner crystal by interpolating between our exact classical and quantum few-electron calculations.
 
  The rest of this paper is organized as follows.  In Sec.~\ref{limits}, we describe our theory for effective finite-temperature 1D classical Wigner crystals by calculating the exact Boltzmann distribution function and the theory for the corresponding $T=0$ quantum ground state by using the exact matrix diagonalization technique. In Sec.~\ref{phonon}, we provide a smooth interpolation between the quantum and classical regimes, considering the exact phonon modes of the Wigner crystal and using a generalized Lindemann criterion.  We conclude in Sec.~\ref{conclude} with a summary.
  
  \section{High-temperature and zero-temperature limits}\label{limits}
  
  In this paper, we model a system of $N$ Coulomb-interacting electrons confined in a trapping potential by the Hamiltonian
  \begin{equation}\label{Hamiltonian}
  \begin{split}
     H&=\frac{\hbar^2}{m}\sum_{i=1}^{N}\left[\frac{-\partial^2}{2\partial x_i^2} +  \frac{1}{L_0^2}\left(\frac{2x_i}{L_0}\right)^p\right] \\
     &\quad +\sum_{i< j} \frac{\hbar^2}{ma_B} \frac{1}{\sqrt{(x_i-x_j)^2+d^2}},
  \end{split}
  \end{equation}
  where $p=4$ for consistency with the quartic potential used in Ref. \cite{exp0}, but in principle $p$ can be any even integer; $L_0$ is length scale of the trapping potential; $m$ and $a_B$ are the effective electron mass and Bhor radius. We emphasize that $d$ in our context is a generic parameter which determines the short-distance cut-off for the electron-electron Coulomb interaction but can also be loosely interpreted as the transverse size of the system in experimental contexts. For softer confinement, as in semiconductor quantum wires, one may have to obtain the effective $d$-value by first solving the transverse quantization problem and then taking the appropriate matrix element of the 3D Coulomb interaction in this transverse basis \cite{transverse1, transverse2}. The trapping potential defines a natural length scale $L_0$ and energy scale $E_0=N^2\hbar^2/(mL_0^2)$. By scaling $x'=x/L_0$ and $H'=N^2H/E_0$, we obtain the dimensionless Hamiltonian
  \begin{equation}\label{Hamiltonian2}
  	   H'=\sum_{i=1}^{N}\left[\frac{-\partial^2}{2\partial x_i^{'2}} + \left(2x'_i\right)^p\right] +\sum_{i< j} \frac{Nr_s}{\sqrt{(x'_i-x'_j)^2+\eta^2}}.
  \end{equation}
 The system profile is thus tuned by two dimensionless parameters
  \begin{equation}
  	r_s=\frac{L_0}{Na_B} \text{ and } \eta=\frac{d}{L_0}.
  \end{equation}
  Accordingly, the typical Coulomb interaction energy is $E_c = r_sE_0$. In the following sections, we conduct the numerical simulations for different pairs of $r_s$ and $\eta$ at the classical high-$T$ limit and the quantum $T=0$ limit.
  
  \subsection{Classical high$-T$ limit}
\begin{figure}
  \begin{minipage}{0.2cm}
  	  \rotatebox{90}{\centering $\rho(x)L_0$}
  \end{minipage}
  \begin{minipage}{0.42\textwidth}
  	  \centering
  	  \includegraphics[width=0.95\textwidth]{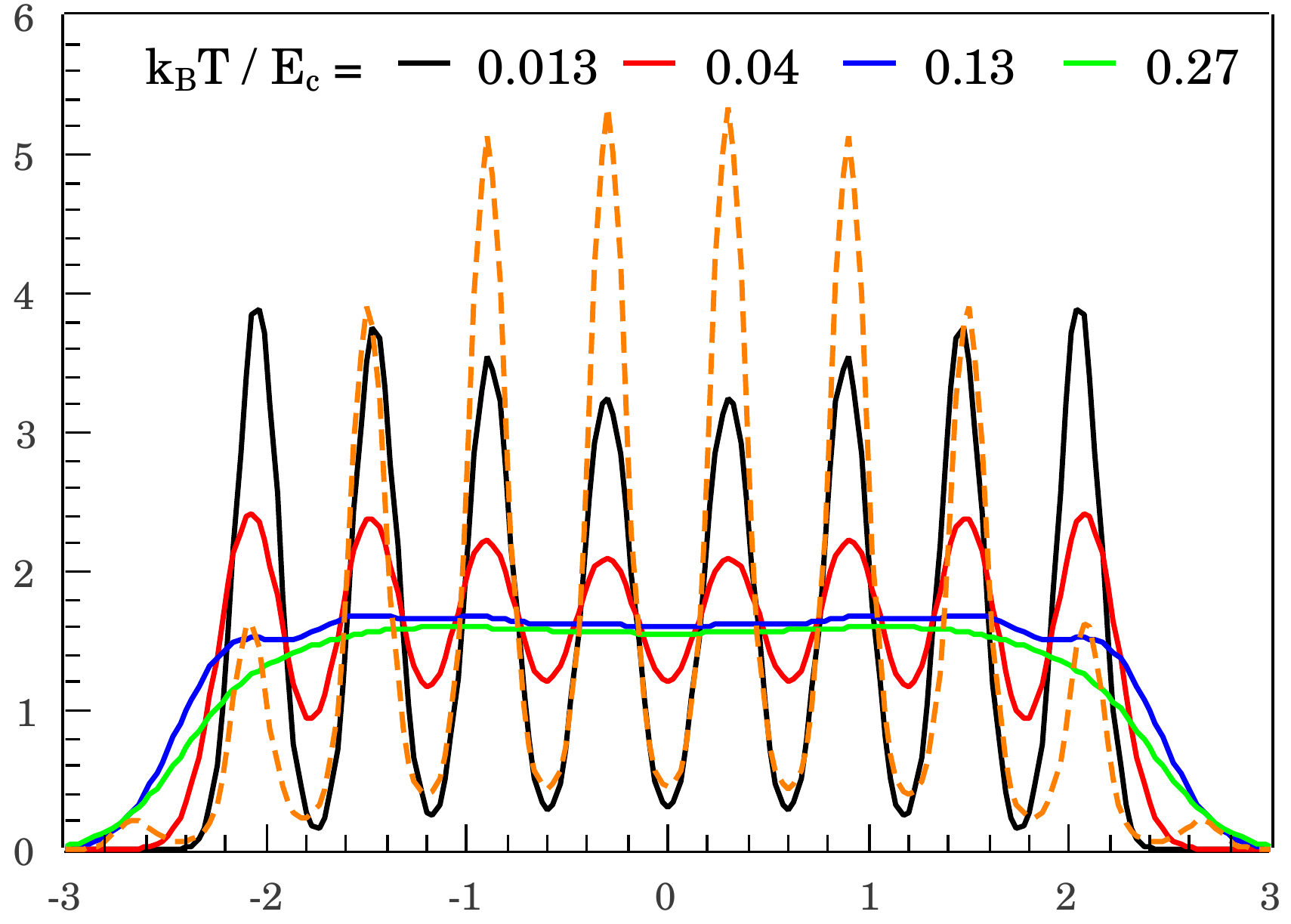}
  	  \hspace*{0.8cm} $x/L_0$
  \end{minipage}
  \hspace*{0.5cm}
  \caption{Spatial density profile for $N=8$ interacting spinless electrons at various temperatures at fixed $r_s=15$ and $\eta=0.01$ (solid lines).  The peak structure emerges preeminently as temperature decreases implying stronger crystallization at lower temperatures. For $k_BT > 0.3 E_c$, the interaction-induced density variation essentially vanishes. If the trapping potential is periodically modulated (see text), the peak pattern in the spatial density profile can appear even without interaction (dashed line).}	\label{fig1}
  \end{figure}

\begin{figure}
  \begin{minipage}{0.2cm}
  	\rotatebox{90}{\hspace*{0.4cm} $C(\Delta x)L_0$}
  \end{minipage}
  \begin{minipage}{0.42\textwidth}
  	\centering
  	\includegraphics[width=0.95\textwidth]{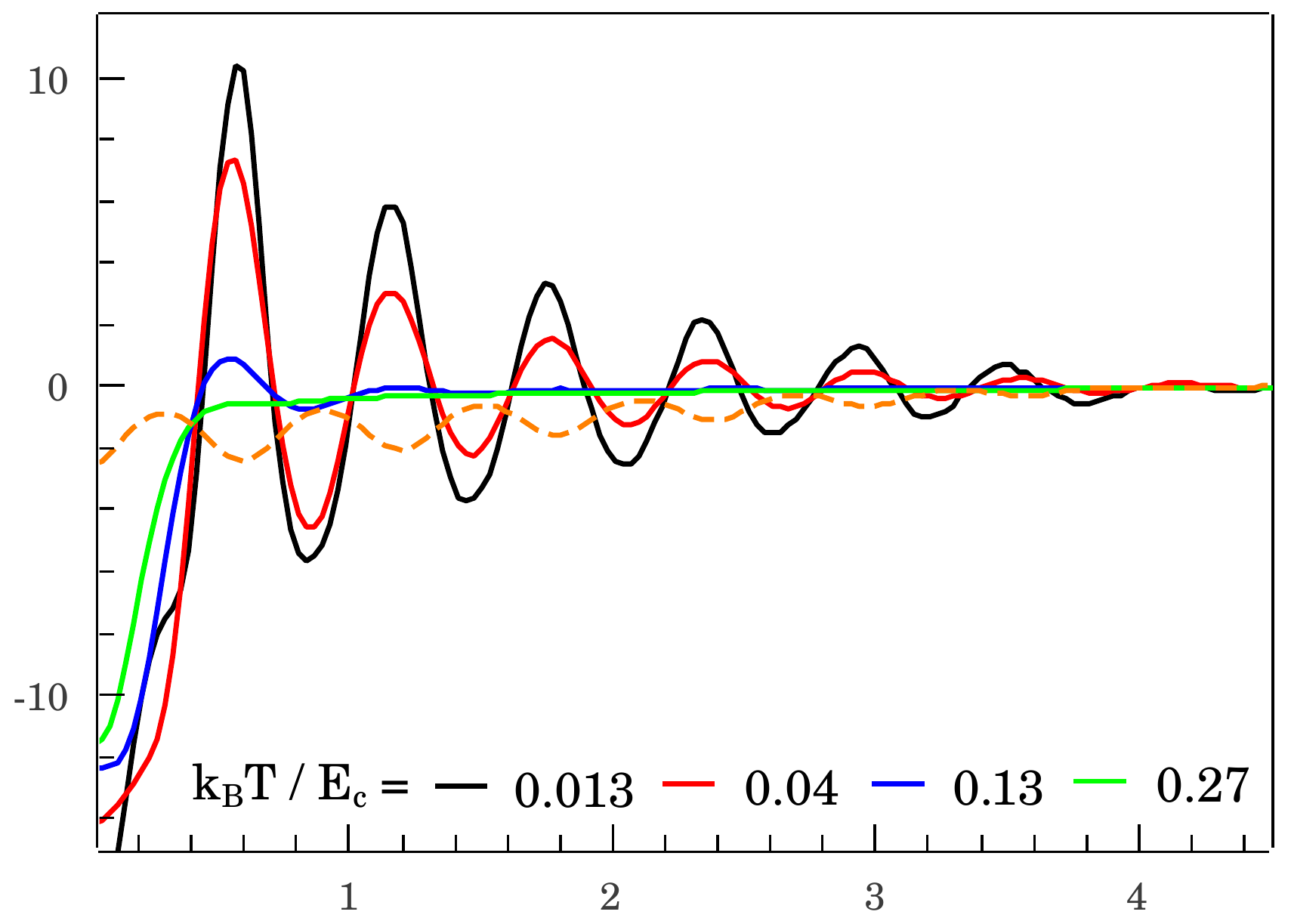}
  	\hspace*{0.3cm} $\Delta x/L_0$
  \end{minipage}
  \hspace*{0.5cm}
    	\caption{Average density-density correlation functions versus the distance for the same parameters as in Fig.~\ref{fig1}. The correlation persists throughout the finite system for low temperatures, but essentially suppressed at high temperatures. Note that for the non-interacting system, the correlation function is always negative despite the possible oscillatory pattern induced by an underlying lattice potential.}	\label{fig2}
    \end{figure}

  The spatial density profile in the classical limit is obtained from the Boltzmann distribution
  \begin{equation}\label{mc}
  \rho(x) \propto \int \delta(x'_1-x)\exp\left(-\frac{U(\{x'_i\})}{k_BT}\right)dx'^N,
  \end{equation}
  where $U$ is the sum of the trapping potential and the interaction energy of the Hamiltonian \eqref{Hamiltonian2}. We evaluate integral \eqref{mc} with the uniform-sampling Monte Carlo method and obtain the spatial density profile at various temperatures. We consider $N=8$ spinless electrons for our calculation as shown in Figs.~\ref{fig1} and \ref{fig2}. In addition, we consider a case of non-interacting electrons subjected to an underlying lattice potential $V_l = 4E_0\cos(10\pi x/3L_0)$ where the density modulation is entirely due to this periodic potential.
  
  The noteworthy feature of the results in Fig.~\ref{fig1} is that the density variation associated with the effective 1D classical Coulomb crystal is  suppressed for $k_BT \gtrsim E_c$ even though for a small range of temperature the variation can be temporarily enhanced \cite{thermal_enhance1,thermal_enhance2,wignermol}. Moreover, we show in Fig.~\ref{fig1} that an underlying periodic potential can potentially produce a similar density variation pattern even in a non-interacting system. This pattern, however, is not the anticipated Wigner oscillation because it is not driven by the Coulomb interaction. Thus, the spatial density is not always a good indicator of the Wigner crystal, particularly if there is a background periodic potential in the system. For a more conclusive evidence, we propose the measurement of the average density-density correlation
  \begin{equation}\label{correlation}
  	  	C(\Delta x)=\int \left[\braket{\rho(x)\rho(x+\Delta x)} -\braket{\rho(x)}\braket{\rho(x+\Delta x)}\right]dx.
  \end{equation}
  For a non-interacting system, the probability of finding $N$ (distinguishable) electrons at \{$x_1,x_2, \dots ,x_n$\} is $P(\{x_1,x_2,\dots, x_n\}) = P_s(x_1)P_s(x_2)\dots P_s(x_N)$ with $P_s$ being the single-particle distribution satisfying $\int P_s(x)dx=1$. Thus $\braket{\rho(x)\rho(y)} = N(N-1)P_s(x)P_s(y)$, while $\braket{\rho(x)}\braket{\rho(y)} = N^2P_s(x)P_s(y)$. As a result, the correlation $C(\Delta x)$ is always negative for a non-interacting system independent of how oscillatory its density distribution might be. In Fig.~\ref{fig2}, we show the calculated density-density correlation with respect to the distance $\Delta x$. For the Coulomb interacting system at different temperatures (solid lines), the modulation is also suppressed by high temperature and disappears at the same temperature as the modulation in the density profile, so the density-density correlation does not provide extra information in this case. However, for the non-interacting system with a periodic underlying potential, the correlation does show an oscillatory pattern like its density profile, but it is always negative unlike the Coulomb-driven Wigner oscillation. Therefore, the sign of the correlation can be used to distinguish whether the density modulation is caused by the interaction or by external potentials. We emphasize that the Coulomb-driven density-density correlations must vanish at long distances even for the classical system, as shown by Peierls a long time ago, because the Debye-Waller factor always diverges in 1D indicating the absence of a true 1D crystal. In a finite system, however, the long-distance thermal disordering of the crystalline order does not manifest itself at sufficiently low temperatures as our results show explicitly.

   \subsection{Quantum limit for spinless system}
   
   \begin{figure}[ht!]
	\begin{minipage}{0.5cm}
		\rotatebox{90}{\hspace*{0.3cm} $\rho(x)L_0$}
	\end{minipage}
	\begin{minipage}{0.42\textwidth}
		\centering
		\includegraphics[width=\textwidth]{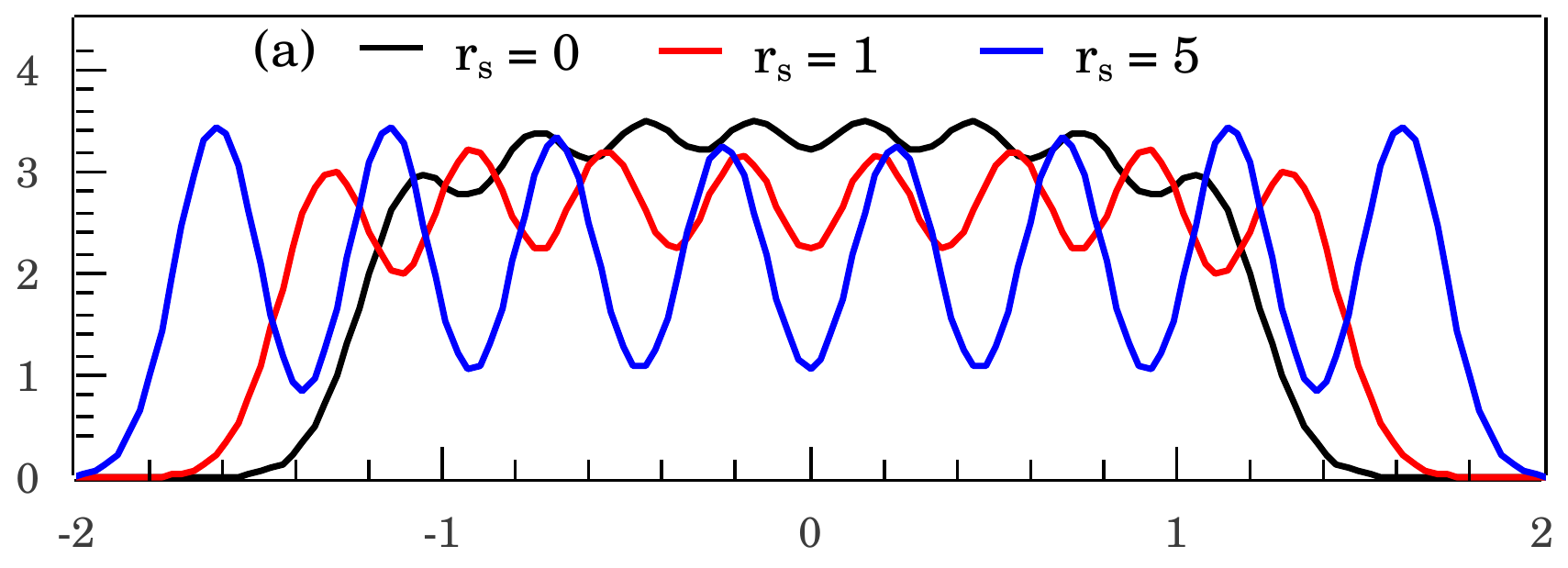}  	
	\end{minipage}
	$x/L_0$
	
	\begin{minipage}{0.5cm}
		\rotatebox{90}{\hspace*{0.3cm} $C(\Delta x)L_0$}
	\end{minipage}
	\begin{minipage}{0.42\textwidth}
		\centering
		\includegraphics[width=\textwidth]{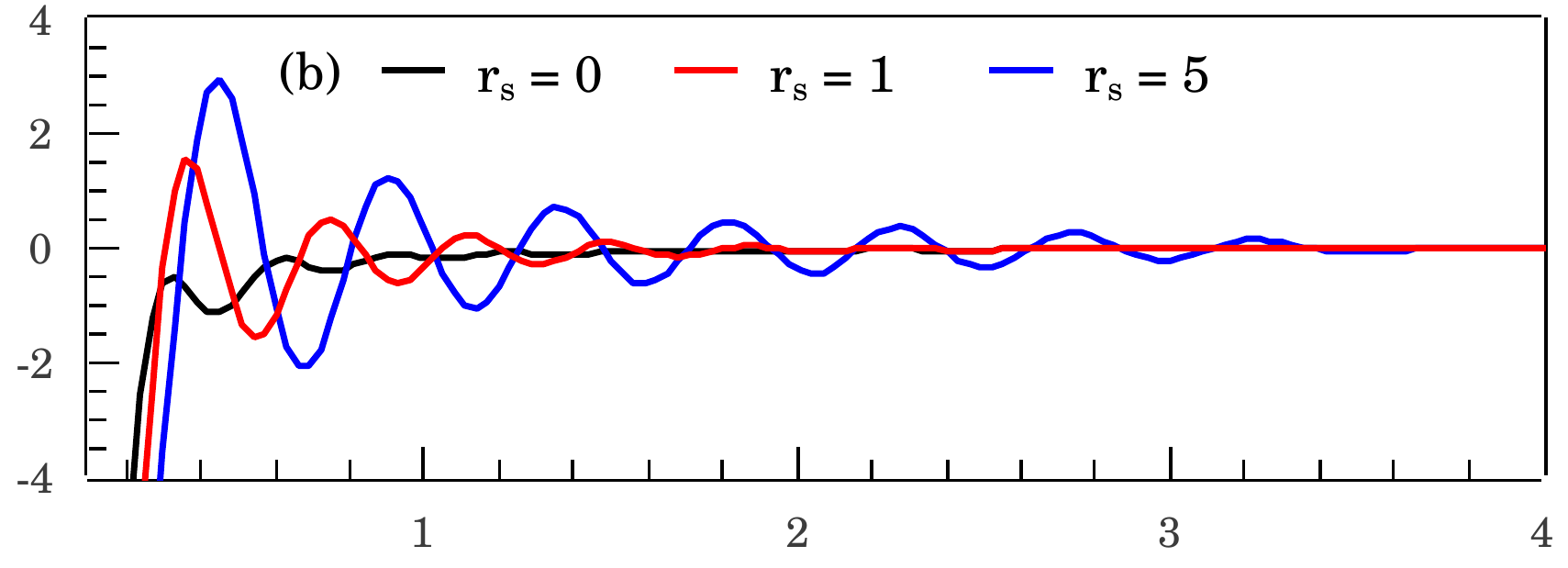}  	
	\end{minipage}
	$\Delta x/L_0$	
	
	\caption{(a) Spatial density profile of a spinless 8-electron system at fixed $\eta=0.01$ and increasing $r_s$. Even for the non-interacting case $r_s=0$, the $N-$peak pattern is still visible. (b) Density-density correlation as a function of distance $\Delta x/L_0$. The correlation function of the non-interacting case is also oscillating but always negative.}\label{fig3}
\end{figure} 

   In this section and the following ones, we study the quantum ground state at $T=0$ by performing exact diagonalization for a small electron system. We use the configuration interaction method, where each configuration is described by a Slater determinant built from single-particle solutions of the free Hamiltonian. We use up to 25 single-particle wavefunctions and keep up to 20000 determinants having the lowest energy. We check for and ensure the convergence in each case studied here.
 	
A spinless system can be realized by a system that is strongly polarized or has infinite on-site interaction \cite{coulomb1}. Spinless particles tend to be apart from each other even without Coulomb repulsion by virtue of the Pauli principle. As shown in Fig.~\ref{fig3}, for the non-interacting system ($r_s=0$), the calculated quantum spatial charge density profile still shares the same number of peaks as the corresponding interacting cases. The non-interacting spinless (or more generally infinite zero-range interaction in a spinful system \cite{noorder}) peaks  arise simply from progressively filling up the bound states of the trapping potential with one electron per energy level.  However, the long-range Coulomb interaction enhances the contrast between these peaks and separates them spatially in order to minimize the Coulomb repulsive potential, thus making the system's spatial density profile similar to that of an effective crystal. Specifically, the contrast defined by the ratio between the variation amplitude and the average density takes the values of 4\%, 18\% and 53\% for $r_s=0,1$ and 5 respectively. Thus, although both noninteracting and interacting situations manifest $N$ density peaks, there is a quantitative difference in the strength of these peaks between the two cases. At larger $r_s$ (lower average density), we expect the Coulomb potential to dominate over the kinetic energy, leading to a progressively more distinct crystalline-looking structure. 

Previous studies try to distinguish between the non-interacting and interacting $N-$peak pattern by computing the two-particle density and find that this function is smooth for a non-interacting system but displays strong modulation for a Coulombic system \cite{noorder,wignercorrelation}. We note that this distinction is purely quantitative because the two-particle density correlator of a non-interacting system does have a small oscillatory part due to the Pauli exclusion principle.  However, by considering the correlation function as in Eq.~\eqref{correlation}, we find a qualitative difference. For a non-interacting system, the many-body wavefunction is expressed by a single Slater determinant $\Psi(x_1,x_2, \dots ,x_N)=\text{Det}M/\sqrt{N!}$ where $M_{ij}=\psi_i(x_j)$ and $\psi$ is a single-particle wavefunction. As a result,
\begin{equation*}
\braket{\rho(x)\rho(y)} - \braket{\rho(x)}\braket{\rho(y)} = -\left| \sum_{i=1}^N  \psi_i(x)\psi_i(y)\right|^2 \le 0.
\end{equation*}
In Fig.~\ref{fig3}(b), we show the computed correlation function corresponding to parameters in Fig.~\ref{fig3}(a). For a strongly interacting system, the strong oscillation extends throughout the system; while for a non-interacting system, there only exists weak oscillation at small $\Delta x$. This non-interacting correlation function, however, is always negative, thus clearly distinguishing it from the interacting counterpart.

\begin{figure}
	\begin{minipage}{0.5cm}
		\rotatebox{90}{\hspace*{0.3cm} $\rho(x)L_0$}
	\end{minipage}
	\begin{minipage}{0.43\textwidth}
		\centering
		\includegraphics[width=0.49\textwidth]{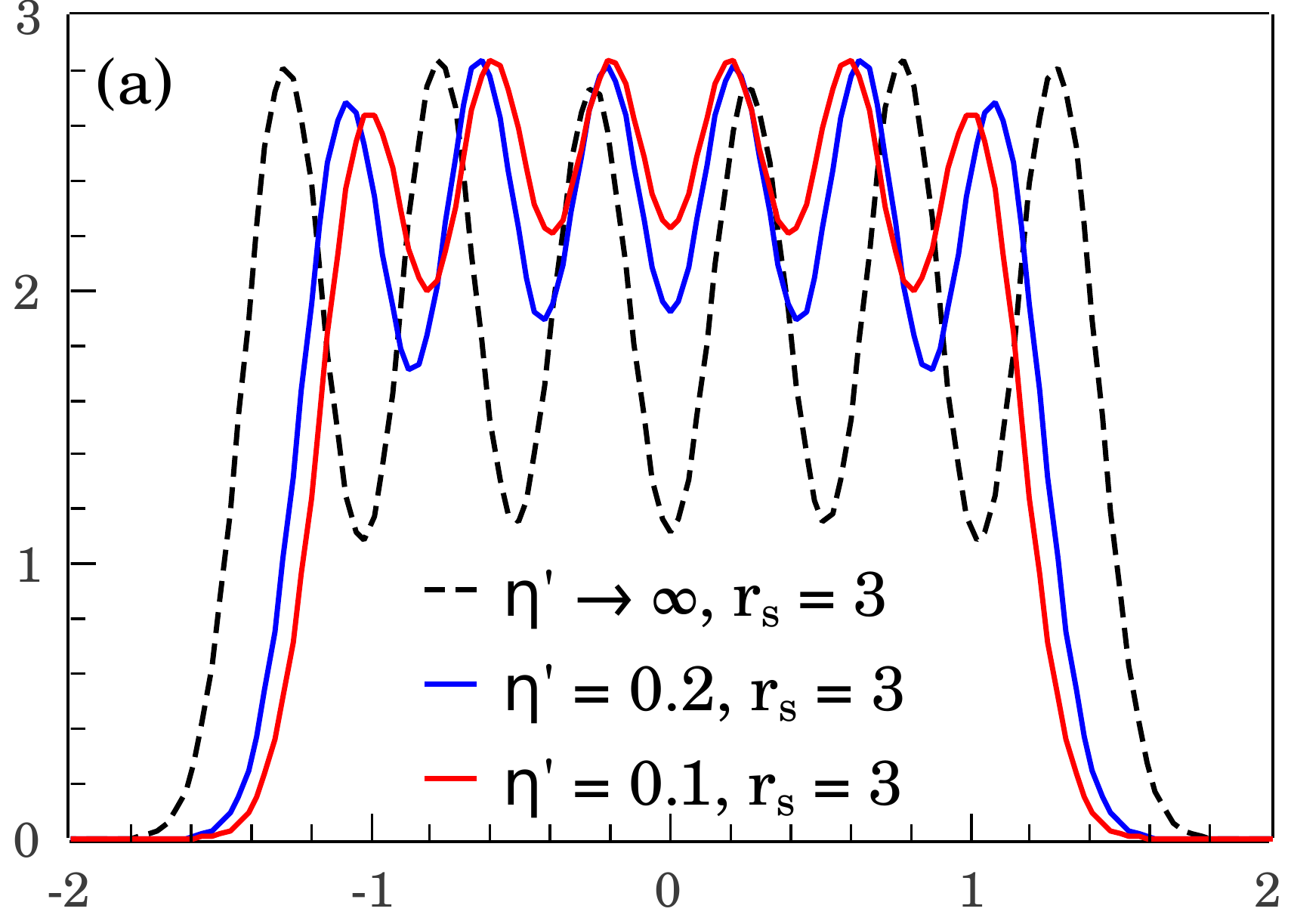}
		\includegraphics[width=0.49\textwidth]{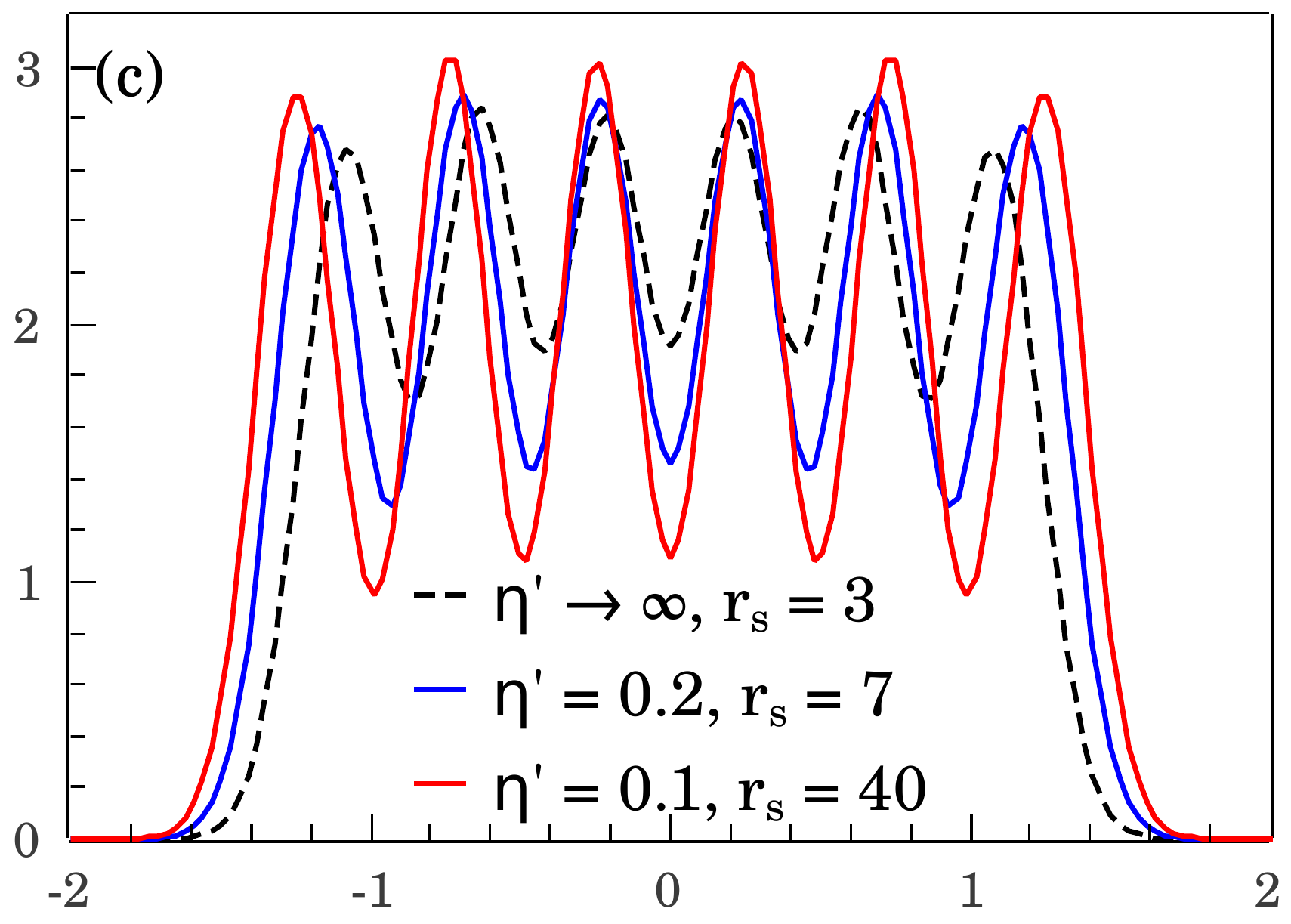}   	
	\end{minipage}
	$x/L_0$
	\vspace{0.1in}
	
	\begin{minipage}{0.5cm}
		\rotatebox{90}{\hspace*{0.3cm} $C(\Delta x)L_0$}
	\end{minipage}
	\begin{minipage}{0.43\textwidth}
		\centering
		\includegraphics[width=0.49\textwidth]{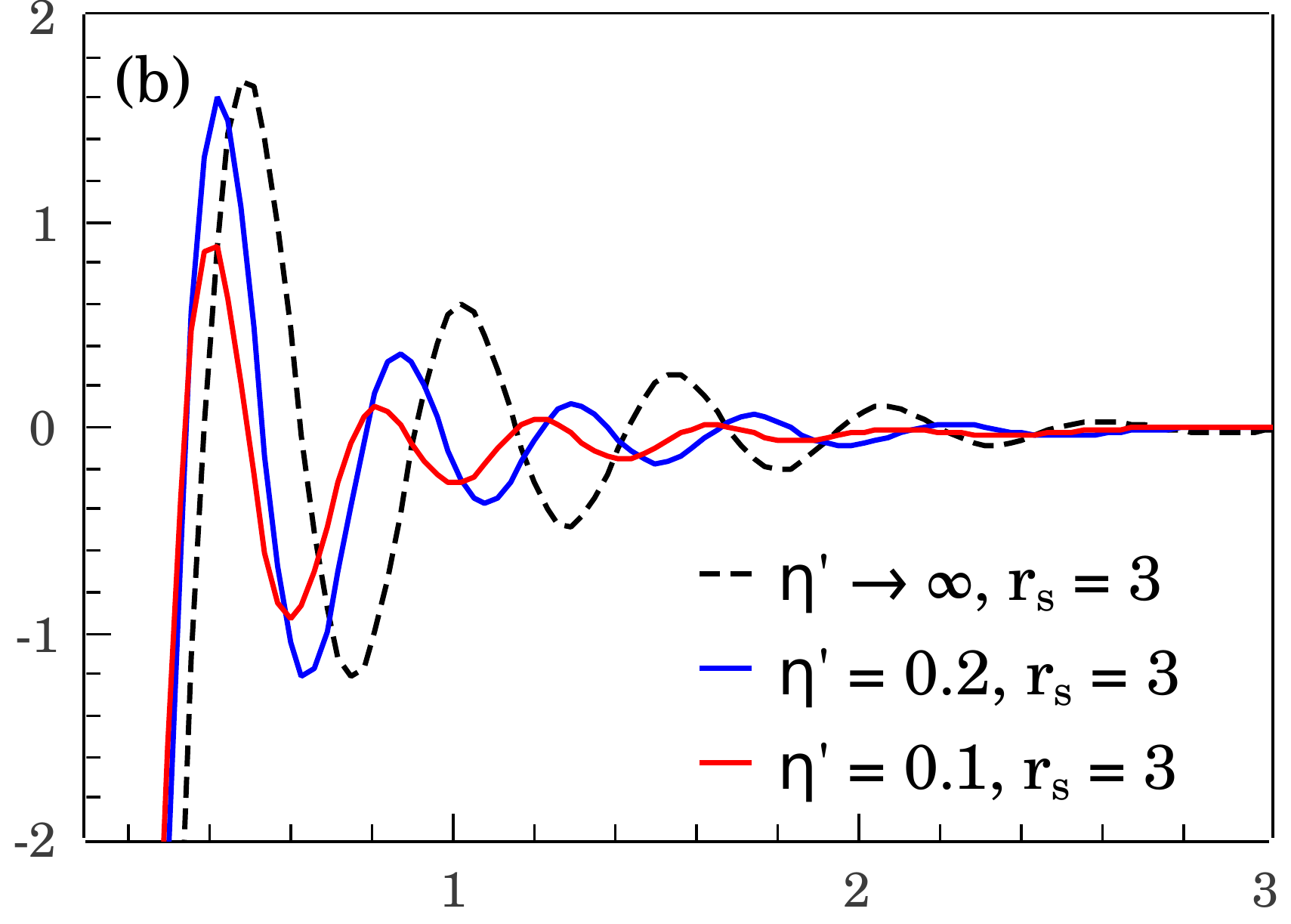}
		\includegraphics[width=0.49\textwidth]{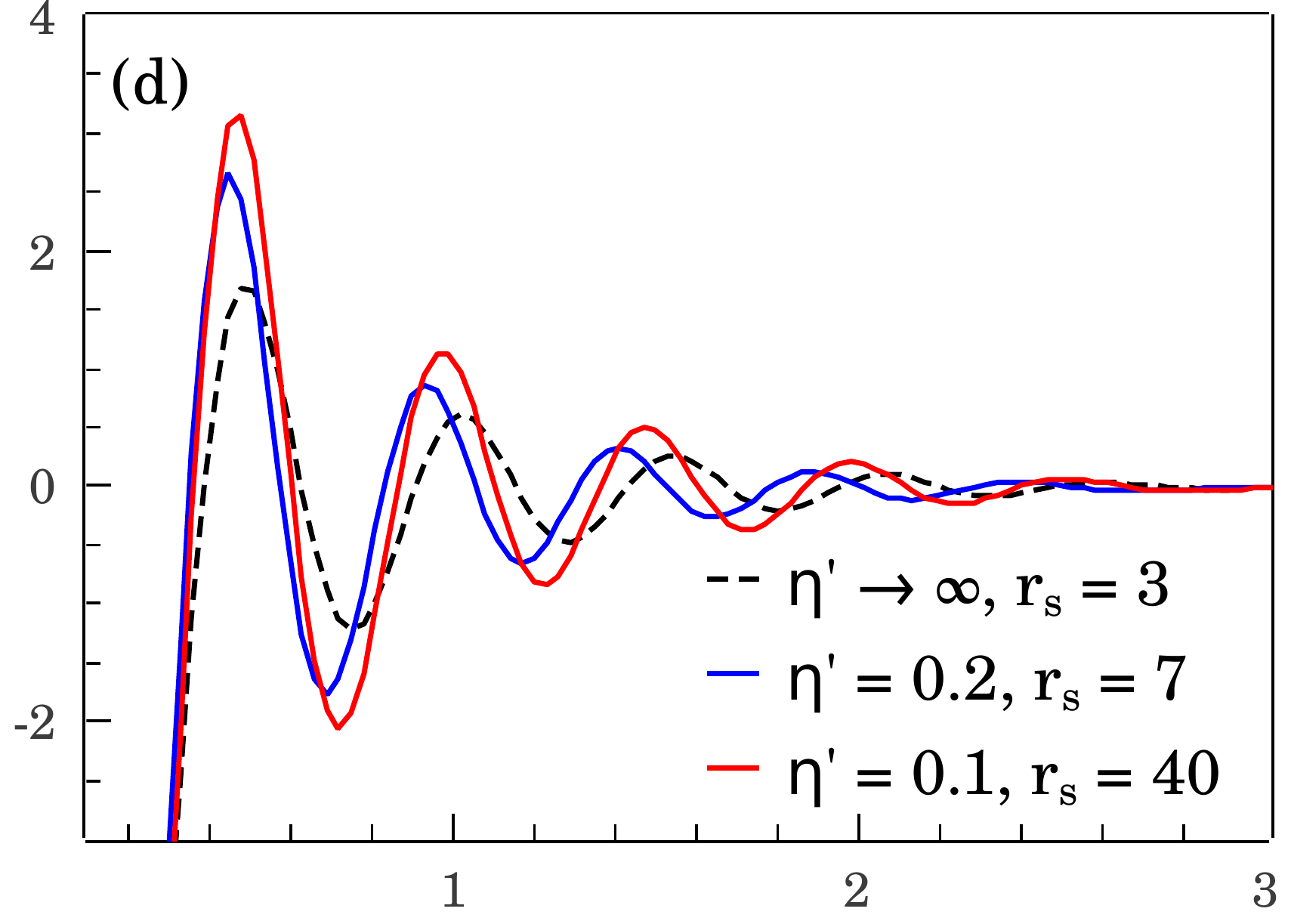}  
	\end{minipage}
	$\Delta x/L_0$	
	
	\caption{(a, c) Spatial density profile of a spinless 6-electron system at fixed $\eta=0.01$. (b, d) Density-density correlation corresponding to Figs (a) and (c)}\label{fig_shortrange}
\end{figure}

From the sign of the correlation function, one can distinguish the non-interacting spinless (or equivalently infinite zero-range interacting spinful system) from the Coulomb interacting system. We then ask whether it is possible to qualitatively differentiate between a finite but short-range interaction and the long-range Coulomb interaction. To study this problem, we repeat the simulation for a system having gated Coulomb interaction $V_g(x') = Nr_s(1/\sqrt{x'^2+\eta^2}-1/\sqrt{x^2+\eta^{'2}})$ which is essentially screened when $x'\gg \eta'$ with $\eta'$ might be tuned by adjusting the distance of the 1D system from a metallic plate. As shown in Fig.~\ref{fig_shortrange}, there is no qualitative difference between the gated and original Coulomb interaction in both density and density-density correlation profiles. More specifically, decreasing $\eta'$ at fixed $r_s$ (see Figs.~\ref{fig_shortrange}(a) and (b)) does suppress the oscillation in both spatial density and correlation profiles. However, by increasing the interaction strength of the screened system (see Figs.~\ref{fig_shortrange}(c) and (d)), the oscillatory pattern can be recovered, making the system a Wigner crystal by any metrics. We then conclude that even though the concept of Wigner crystal is originally for a Coulombic system, any sufficiently strong repulsive non-zero range interaction can form such a crystalline structure in a finite system. The difference between different types of interactions is merely quantitative.

   \subsection{Quantum limit for spinful system}
   \begin{figure*}[ht!]
    \begin{minipage}{0.5cm}
       \rotatebox{90}{\hspace*{0.3cm} $\rho(x)L_0$}
    \end{minipage}	
   	\begin{minipage}{0.92\textwidth}
   		\centering
   		\includegraphics[width=5.4cm]{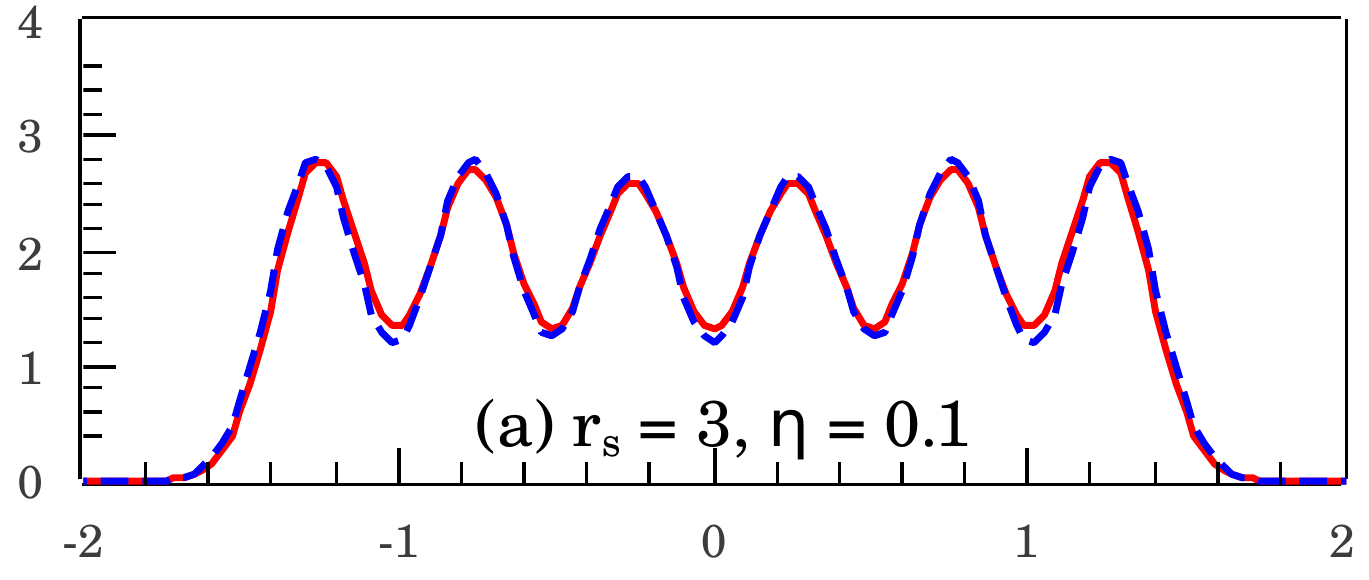}
   		\includegraphics[width=5.4cm]{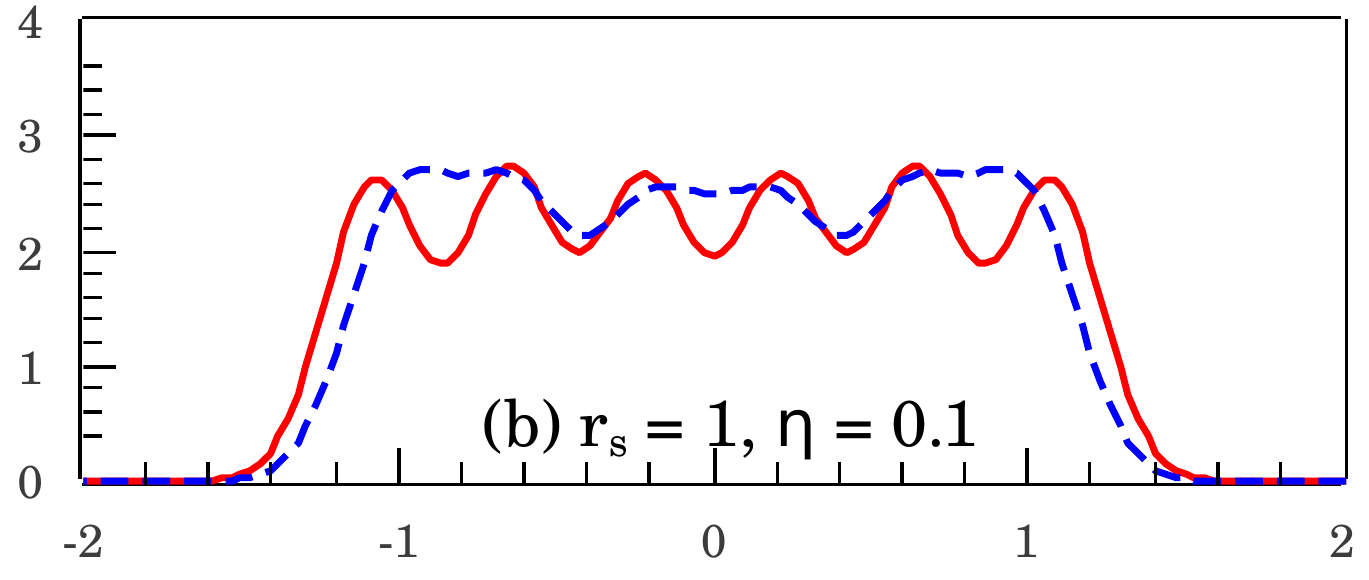}
   		\includegraphics[width=5.4cm]{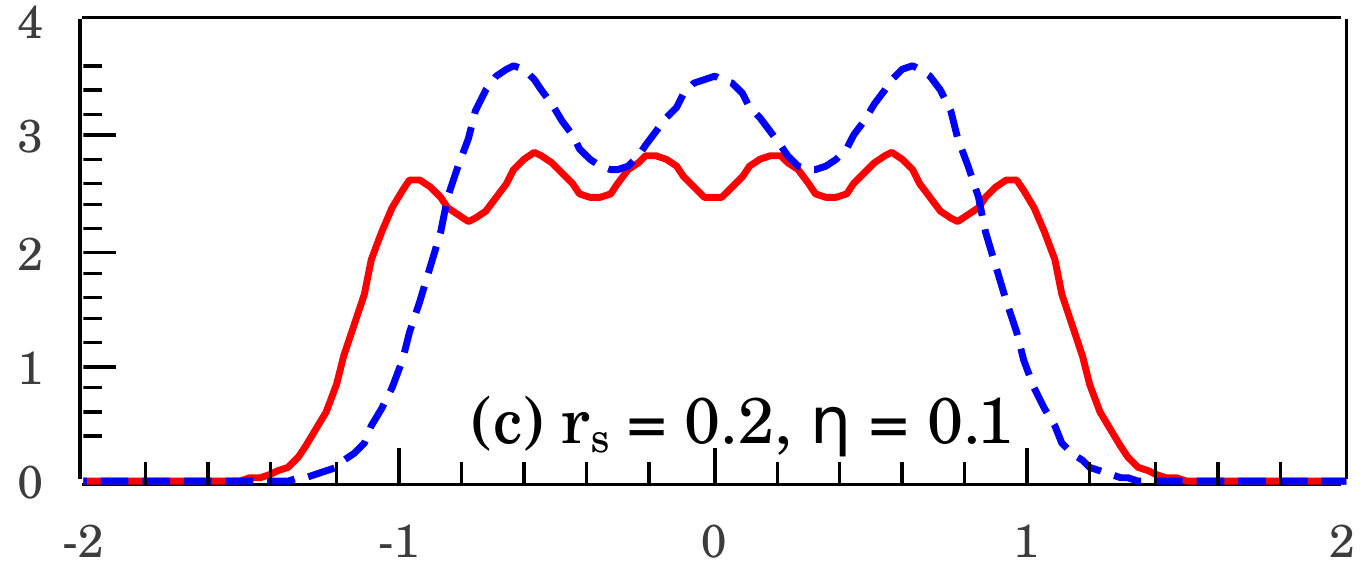}
   		
   		\includegraphics[width=5.4cm]{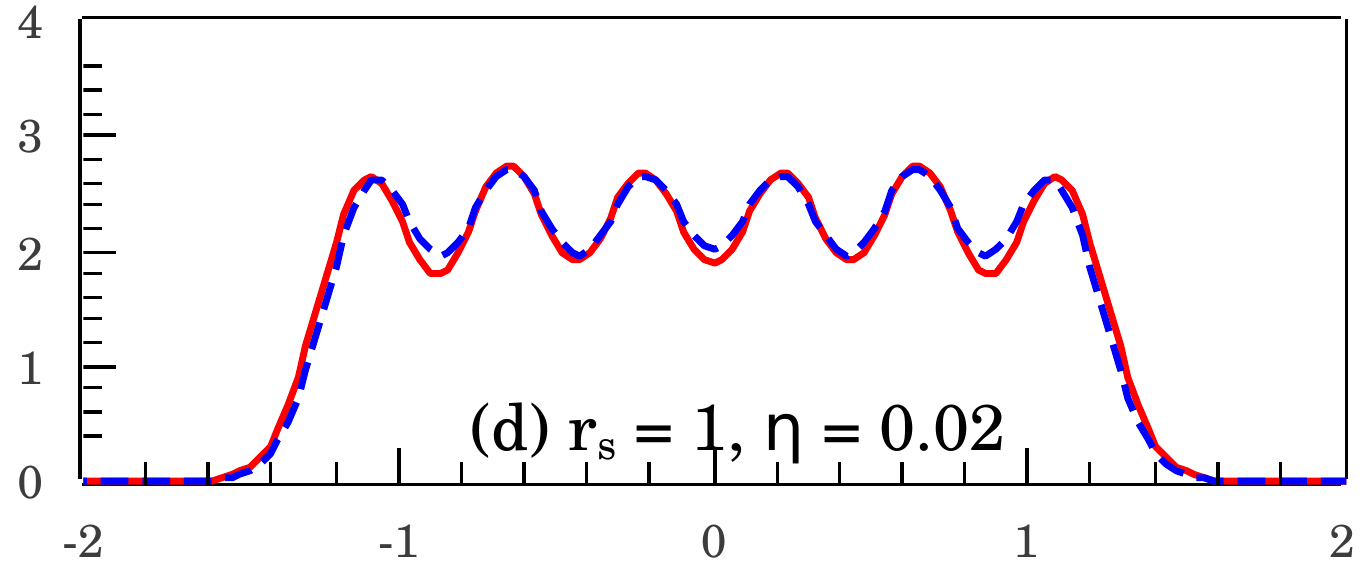}
   		\includegraphics[width=5.4cm]{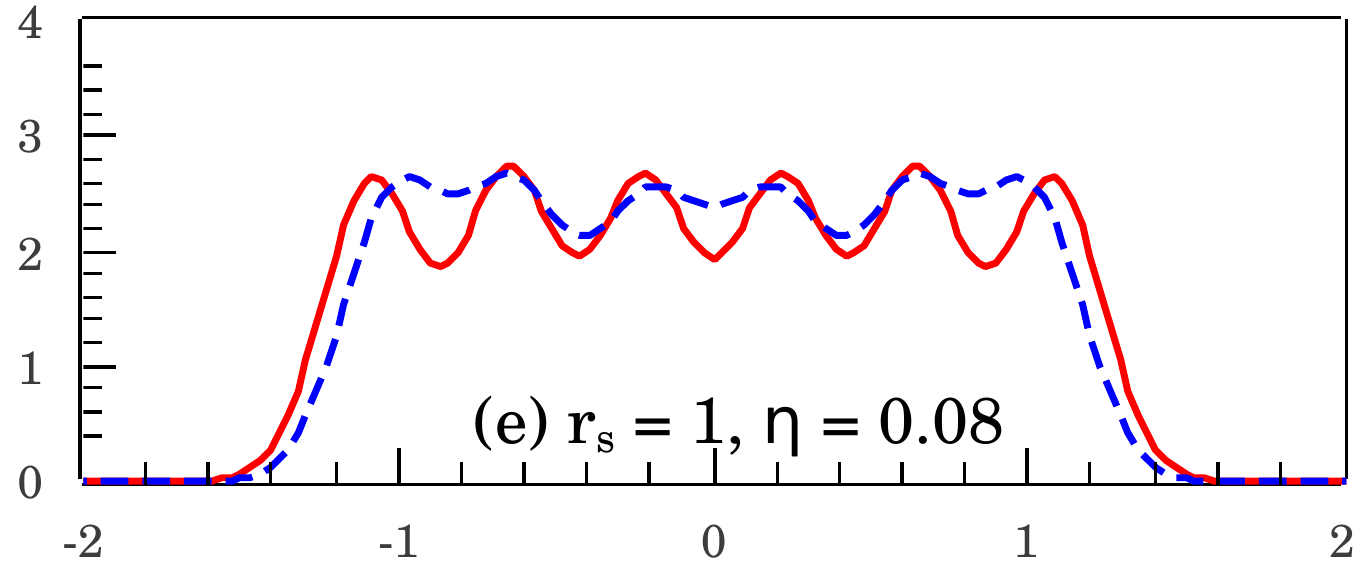}
   		\includegraphics[width=5.4cm]{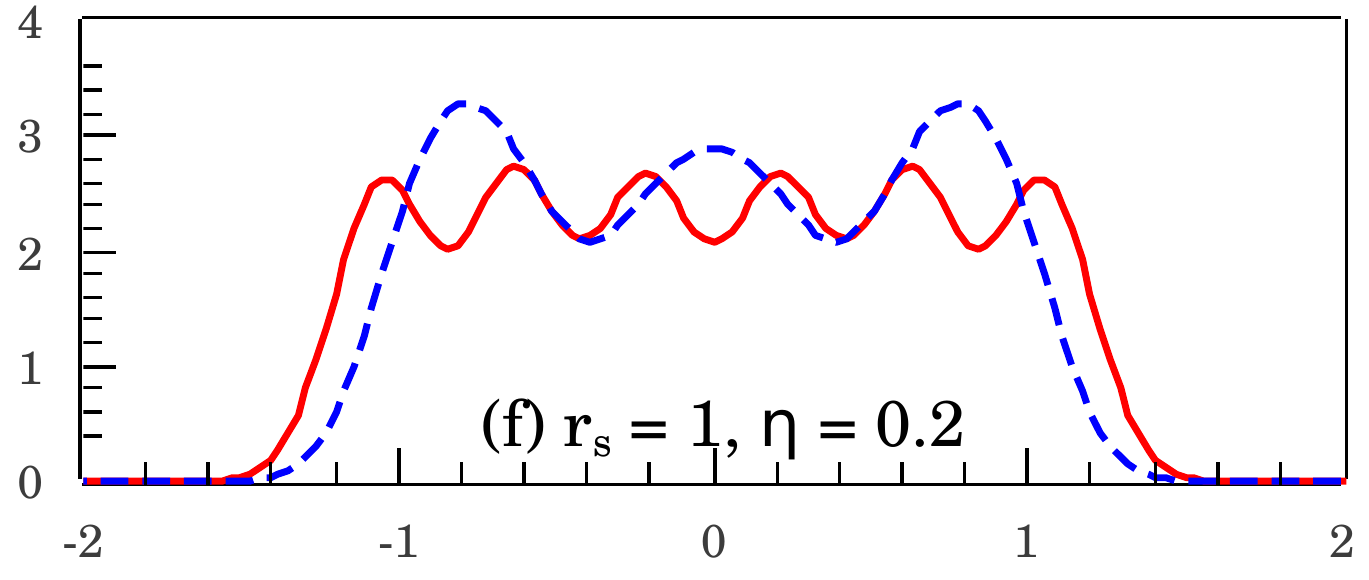}
   	\end{minipage}\\
   	\vspace{0.1in}
   $x/L_0$
   	\caption{(a) Spatial density profile for 1D spinful (blue dash) and spinless (red line) ($N=6$) systems in Wigner crystal phase. The two parameters of the system, i.e. $r_s$ and $\eta$, are also shown. Note that the two systems in the Wigner crystal phase ((a) and (d)) have very similar spatial density distribution. The upper panel: $r_s-$induced phase transition at fixed $\eta$ from (a) $r_s=3.0$ to (b) $r_s=1.0$ and (c) $r_s=0.2$. The lower panel: $\eta-$induced phase transition at fixed $r_s$ from (d) $\eta=0.02$ to (f) $\eta=0.08$ and (e) $\eta=0.2$. The transition manifests as the smear out of spatial density peaks in the spinless system; whereas in the spinful system, the number of peaks is reduced to $N/2$.}	\label{fig4}
   \end{figure*}     
   
   \begin{figure}[ht!]
   	\begin{minipage}{0.5cm}
   		\rotatebox{90}{\hspace*{0.3cm} FFT($\rho(x)L_0$)}
   	\end{minipage}
   	\begin{minipage}{0.44\textwidth}
   		\centering
   		\includegraphics[width=0.49\textwidth]{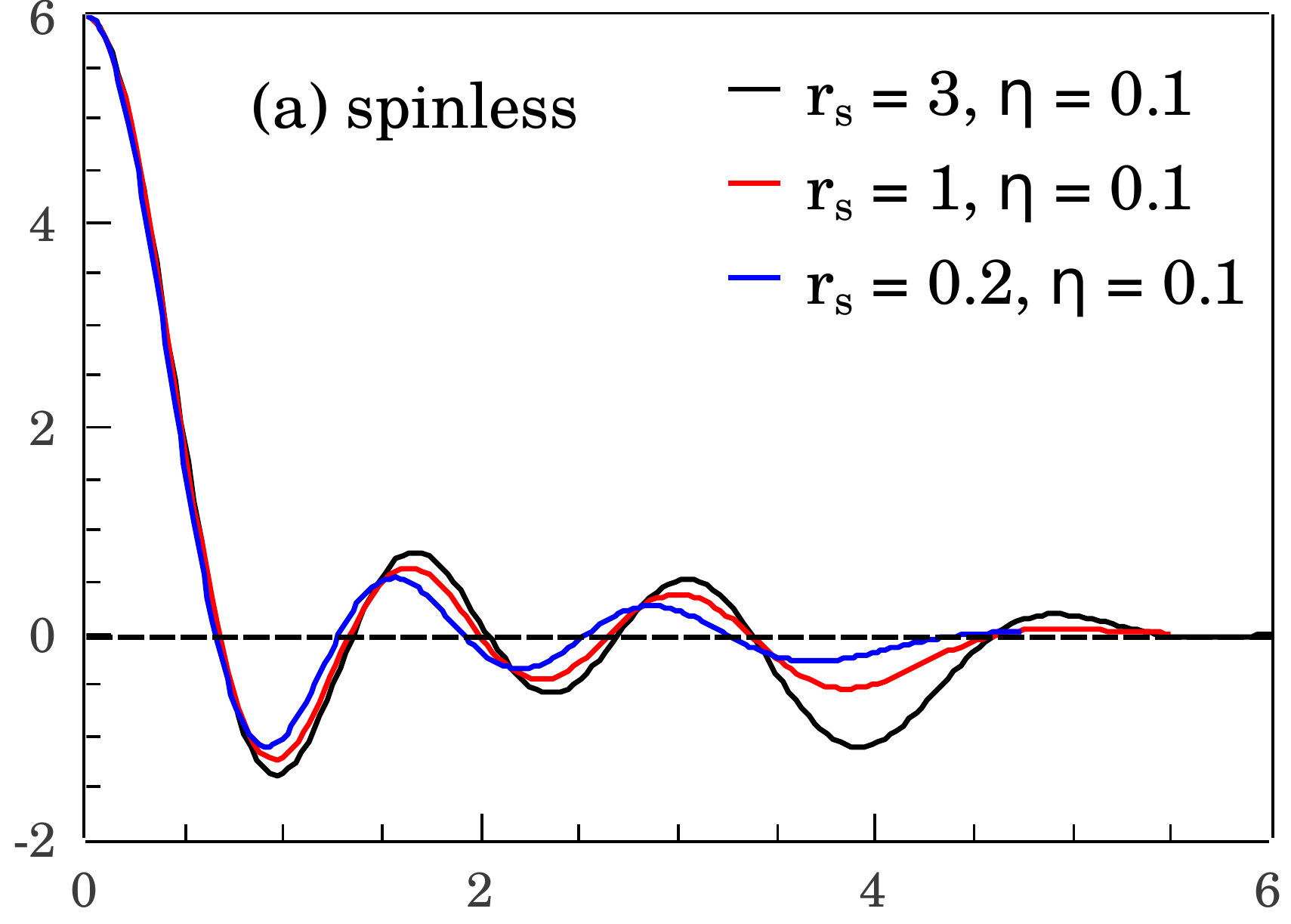}
   		\includegraphics[width=0.49\textwidth]{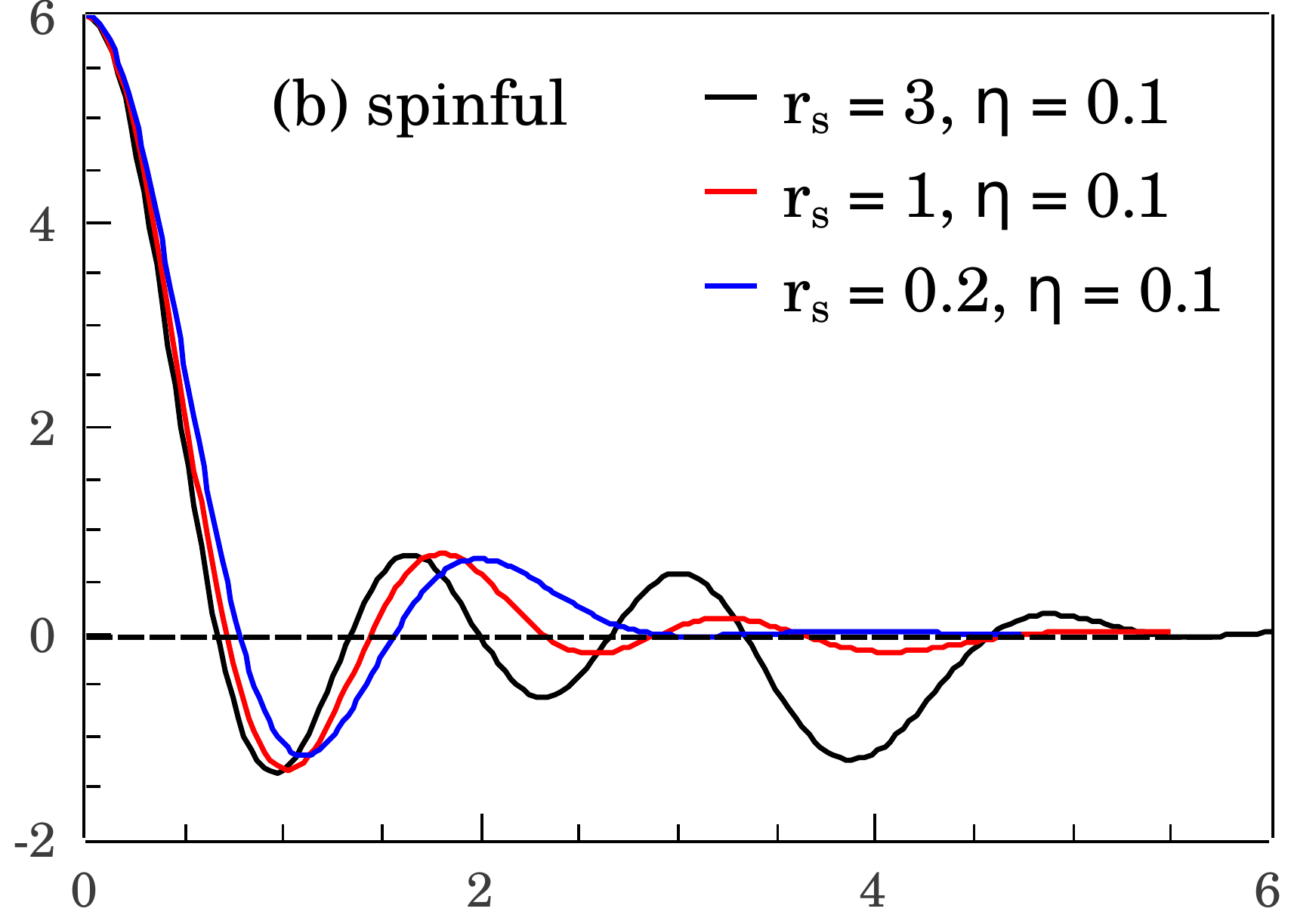}	
   	\end{minipage}
   	$k/k_F$
   	\caption{Fourier transform spinless (a) and spinful (b) systems corresponding to Fig.~\ref{fig4}(a)-(c). The  peak $4k_F$ always exists in the spinless case but is significantly suppressed in the spinful case when the system is deep into the liquid phase.}\label{fig5}
   \end{figure}   
   
   For a spinful system, the system prefers either double occupancy to minimize the kinetic energy or single occupancy to minimize the interaction energy. The 1D solid-liquid crossover is usually defined when the $4k_F$ oscillation becomes visible compared to the usual Friedel $2k_F$ one as the interaction gets stronger. As the average density increases with the electrons coming closer and substantial wavefunction overlap, the kinetic energy term becomes important and doubly filled single-site states may become energetically favorable, inducing a solid-liquid crossover. In Figs.~\ref{fig4}(a), (b), (c), we demonstrate such crossover at fixed $\eta$ induced by decreasing $r_s$ in Eq.~\eqref{Hamiltonian2}. For the spinless system with single-site occupancy, the total number of peaks remains the same and equals the number of electrons. By contrast, the spinful system, starting from the same $N$-peak structure with single-site occupancy for large $r_s$ with negligible exchange effect, eventually manifests only three $(N/2)$ spatial density peaks reflecting double site occupancy for small $r_s$ as the exchange energy becomes  significant. The same physics also applies at fixed coupling parameter $r_s$ but with increasing the cut-off $\eta$, as can be seen in Figs.~\ref{fig4}(d), (e), (f) (see Figs.~\ref{fig10a} and \ref{fig11a} in Appendix B for more simulation results). To better understand the quantitative aspects of the shift in the spatial density oscillatory patterns of the spinful system, we plot the Fourier transform of Fig.~\ref{fig4}(a)-(c) in Fig.~\ref{fig5}. In the spinless case, the peak at $4k_F$  is always enhanced where $k_F=\pi/2a$ is the 1D Fermi momentum in terms of the average spacing $a$. Conversely, this $4k_F$ peak is noticeably suppressed in the spinful liquid phase (albeit being always present in the Coulomb Luttinger liquid). In fact, as emphasized already, the slowly decaying $4k_F$ oscillation is unique to the 1D long-range interacting system and has a much slower spatial decay rate  compared with the $2k_F$ oscillation \cite{Schulz}. However, for a finite system, the competition between these two oscillations is also determined by the system size and the details of the mutual interaction (i.e. the value of $\eta$), leading to the non-universal existence of an effective finite-size 1D Wigner crystal although there is no such solid phase in the infinite 1D system.
   
   Figures~\ref{fig4} and \ref{fig5} suggest that the ratio $r_s/\eta$ might be important in the single-double occupancy crossover. As $r_s/\eta$ decreases, the cost for two electrons to stay close decreases, thus amplifying the wavefunction overlap and the exchange energy. As a result, the spatial density's oscillatory pattern is changed in the spinful system with a reduction in the number of peaks arising from the double occupancy of sites. Our result is consistent with that of Ref.\cite{coulomb1} using the Hubbard model, in which the ratio of the on-site interaction over the tunneling strength is equivalent to our $r_s/\eta$ - by our definition $r_s/\eta = L_0^2/(Na_B d) = \hbar^2/(ma_B d)(E_0/N)^{-1}$.
   
   \subsection{Exchange energy}
   \begin{figure*}[ht!]
	\rotatebox{90}{\hspace{-1.5cm} $J/E_0$ \hspace{0.8cm} $x/L_0$}
	\begin{minipage}{0.31\textwidth}
		\centering
		$\rho(x)L_0$\\
		\includegraphics[width=\textwidth]{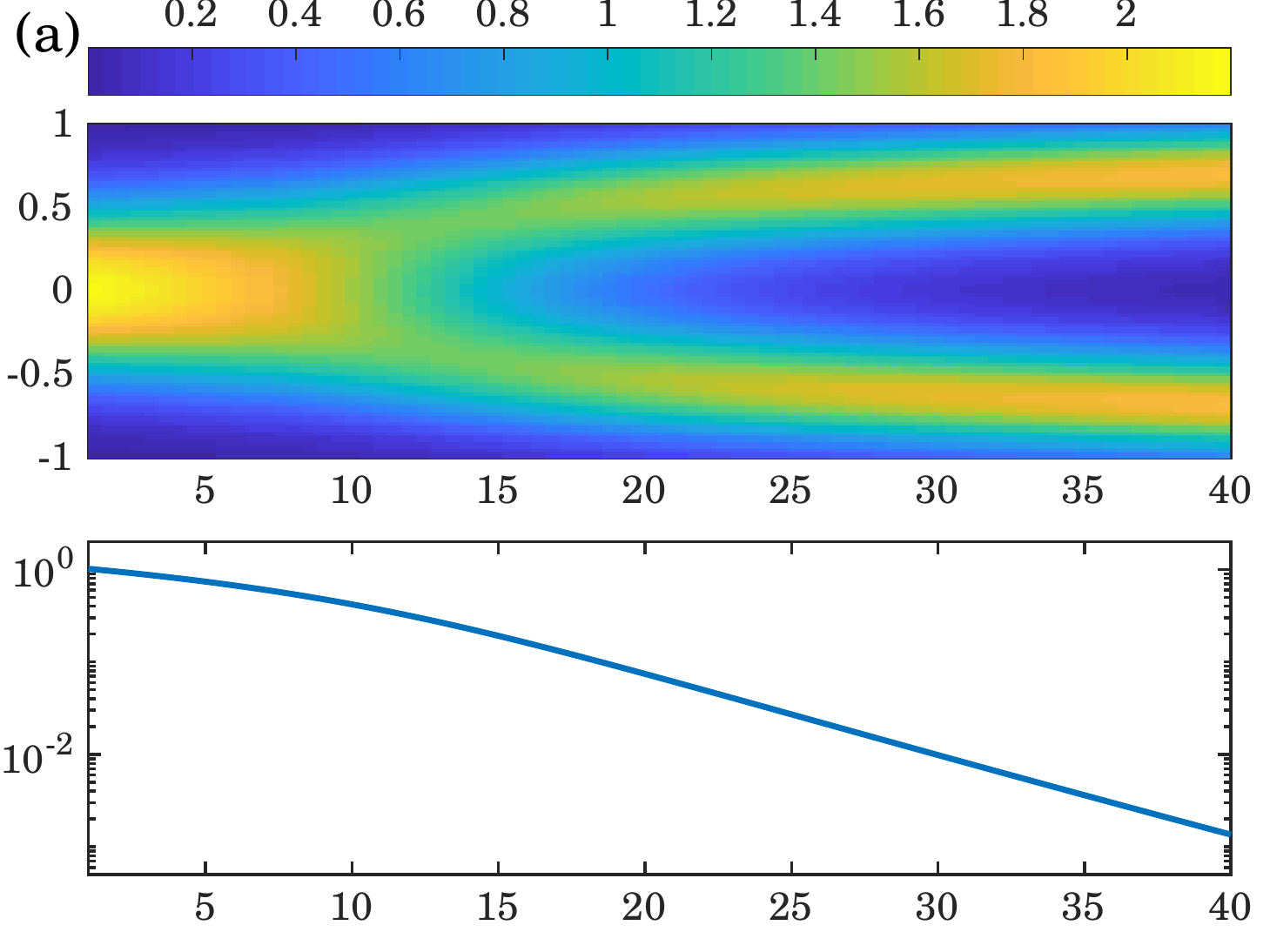}\hspace{1in}
		$r_s$   		
	\end{minipage}
	\begin{minipage}{0.31\textwidth}
		\centering
		$\rho(x)L_0$\\
		\includegraphics[width=\textwidth]{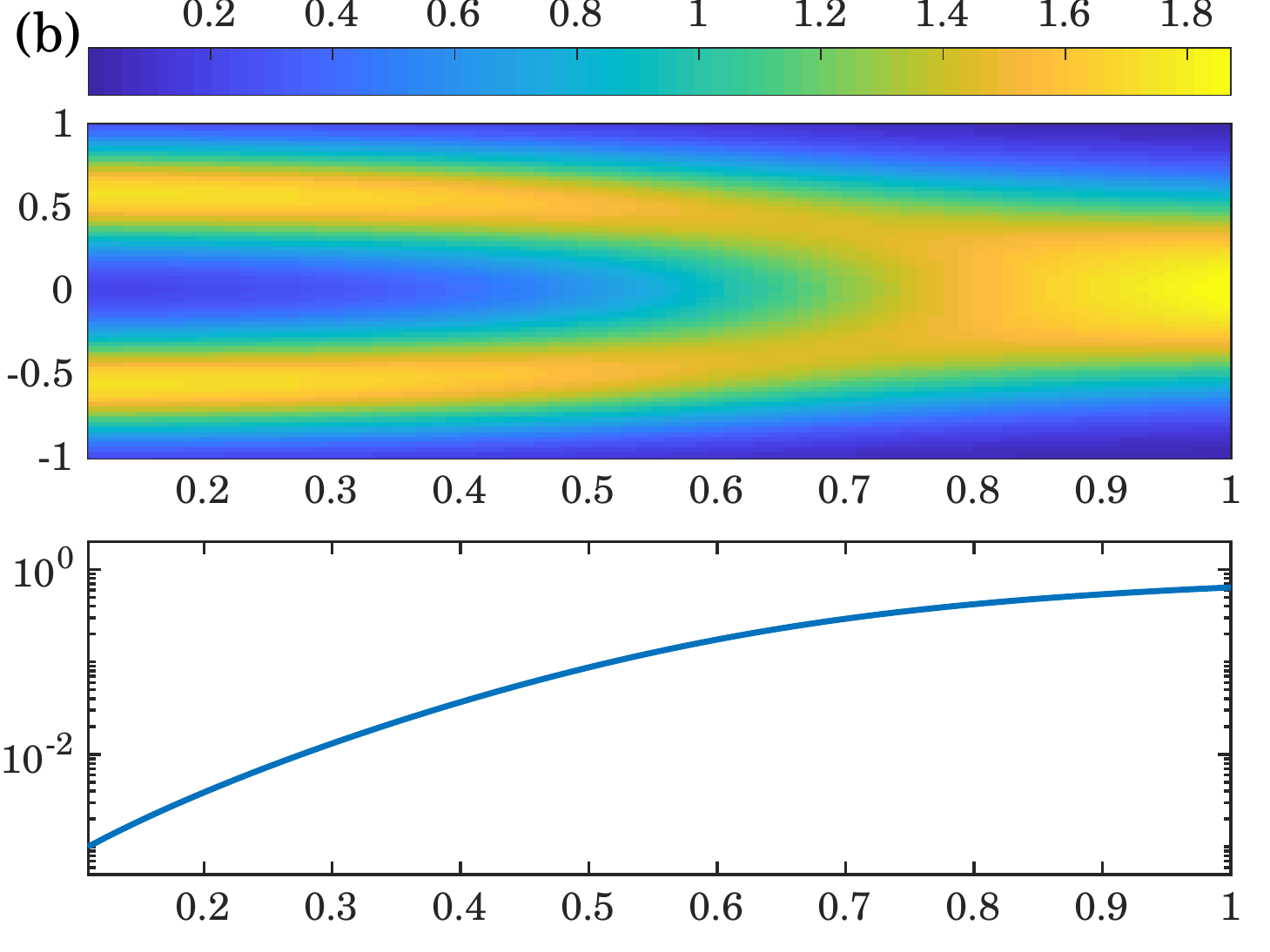}\hspace{1in}
		$\eta$   		
	\end{minipage}   	
	\begin{minipage}{0.31\textwidth}
		\centering
		$\rho(x)L_0$\\
		\includegraphics[width=\textwidth]{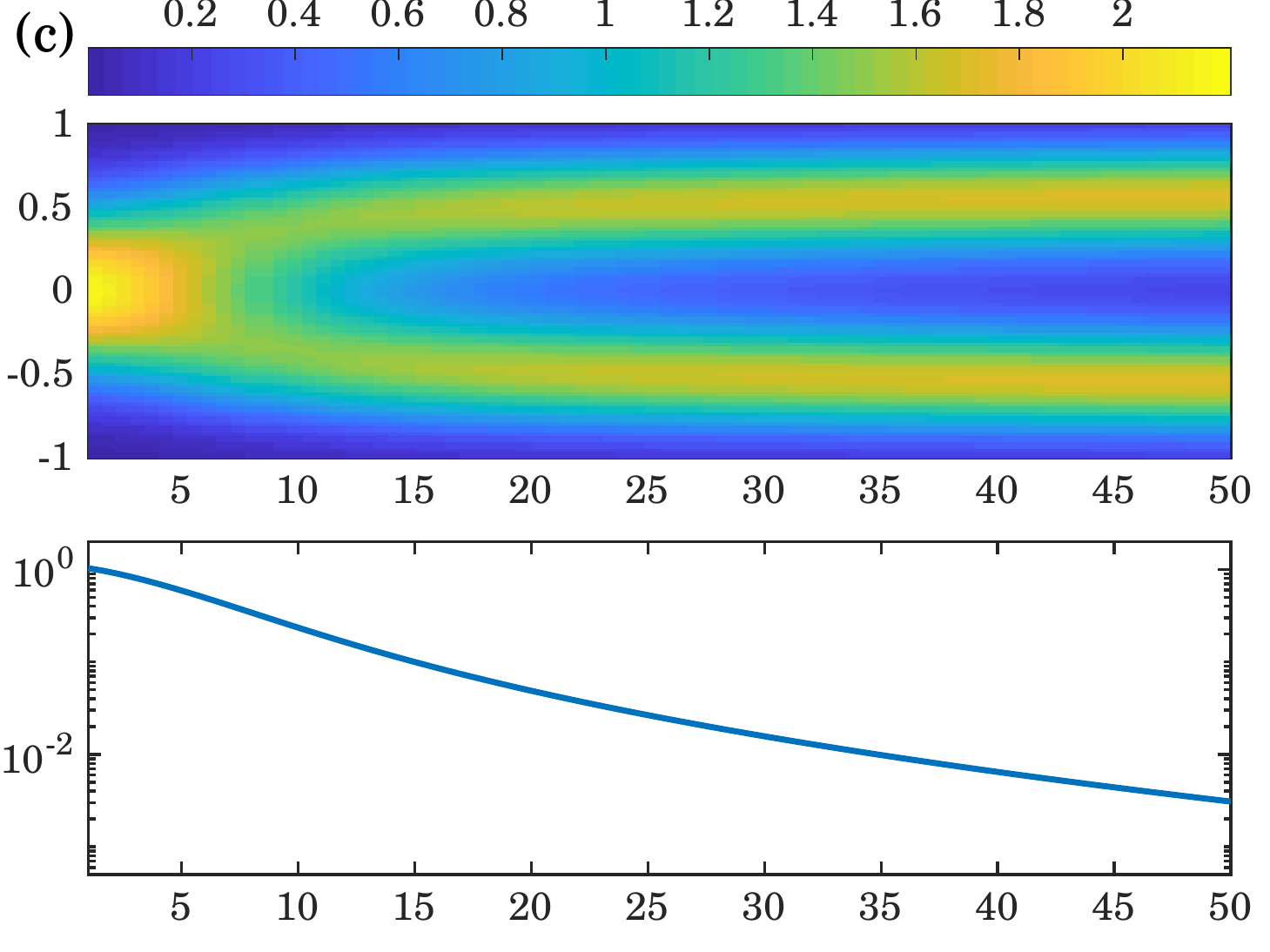}\hspace{1in}
		$r_s/\eta$   		
	\end{minipage}     	
	
	\caption{The spatial density and the relative exchange energy $J/E_0$ in various processes: (a) increasing $r_s$ at fixed $\eta=0.9$, (b) increasing $\eta$ at fixed $r_s=8.0$ and (c) increasing $r_s/\eta$ at fixed $r_s \eta =1.0$. The liquid-solid crossover all happens at around $r_s/\eta\approx 10$. The exchange energy drops to a small value at the same point as the spatial density profile changes between one-peak and two-peak patterns, signaling liquid-solid crossover.}\label{fig11}
\end{figure*}   
   In the effective crystal phase, the localized spatial density peaks are far apart from each other, resulting in a negligible exchange energy. Conversely, in the liquid phase, the overlap increases and the exchange energy becomes significant. In the liquid phase, therefore, spinless and spinful electrons exhibit a qualitative difference. In this subsection, by studying a simple case of 2 particles, we draw a link between the exchange energy and the solid-to-liquid phase transition. Starting with the Hamiltonian \eqref{Hamiltonian2} for $N=2$ particles, we can factorize the wavefunction into the center-of-mass and relative-motion part $\Psi(x_1',x_2')=\phi(x_1'+x_2')\psi(x_1'-x_2')$.  We note that only for a quadratic binding potential ($p=2$) that these two degrees of freedom can be decoupled, otherwise such as $p=4$ in this paper, there are coupling terms appearing from factorizing the binding potential into center-of-mass and relative coordinates. At this point, we ignore these coupling terms and investigate the part of the Hamiltonian containing only $\Delta=(x_1'-x_2')$
   \begin{equation}\label{Eq11}
   	H'_{\Delta}=-\frac{1}{4}\frac{\partial^2}{\partial\Delta^2} + 2\Delta^p +\frac{2r_s}{\sqrt{\Delta^2+\eta^2}}.
   \end{equation}
   For two spinless fermions, $\psi(\Delta)=-\psi(-\Delta)$; while for two spinful fermions, the wavefunction can be symmetric or anti-symmetric depending on the total spin. However, we know that the ground state of 1D single particle has zero nodes and hence the ground state wavefunction $\psi(\Delta)$ must be symmetric for the spinful system (total spin $S=0$). In general, it can be shown that the ground state of a system with an even number of particles is always anti-ferromagnetic; on the other hand, a system with an odd number of particles has the ground state with $S=1/2$ - the smallest possible total spin \cite{singlet}. 
   
   The energy difference between the first excited state (anti-symmetric) of \eqref{Eq11} and the ground state (symmetric) corresponds to the exchange energy between two spinful particles. The potential part of \eqref{Eq11} is a double well potential, thus the energy difference (or the exchange energy) depends on the tunneling through the potential barrier. It is noted that $V(\Delta)>0$ and $V(0)=2r_s/\eta$. Thus, if $r_s/\eta \ll 1$, the barrier is essentially low, leading to large tunneling and large exchange energy. In the other limit, when $r_s/\eta \gg 1$, the barrier is both high and wide, making the exchange energy exponentially small. This shows that the exchange energy, and subsequently the number of density peaks, depends strongly on the ratio $r_s/\eta$ or the short-range behavior of the interaction. This completes the result of \cite{R1} which stated that the long-range part of the interaction does not determine the number of density peaks.

   In Fig.~\ref{fig11}, we show the spatial density profile along with the exchange energy defined as $J=E_{S=1}-E_{S=0}$. The data is obtained numerically from the exact diagonalization during the system transition between the effective solid and liquid phase by varying $r_s$ at fixed $\eta$, varying $\eta$ at fixed $r_s$ and varying $r_s/\eta$ at fixed $r_s\eta$. We emphasize that in all three different schemes, the melting happens around $r_s/\eta\approx 10$, confirming that this ratio is an important factor determining the oscillatory pattern of the Coulombic system. When the two separate peaks start to emerge, signaling the liquid-to-solid crossover, the exchange energy decreases sharply by around 2 orders of magnitude. This shows that in the liquid phase, the singlet state is much more energetically favorable than the triplet state, which is obvious from the previous Wigner molecule argument. Deep in the solid phase, the exchange energy is exponentially small due to large value of $r_s/\eta$. 

  In conclusion, for spinful systems, the increase of the magnitude of the exchange energy reflects the preference of doubly-occupied sites (singlet state) over singly-occupied sites (triplet state), leading to the spinful solid-liquid phase transition. Thus, any liquid-to-solid transition must necessarily accompany a huge decrease in the magnitude of the exchange energy, and the exchange energy in the effective solid phase is exponentially small, being essentially zero for all practical purpose \cite{exp0}. Since the exchange energy in the effective solid phase is likely to be much smaller than the experimental temperature, spin coherence is completely lost in the solid phase, with the thermal spin fluctuations being large. Thus, the effective 1D Wigner crystal is a spin incoherent system at finite temperatures (since the temperature is likely to be much larger than the exponentially small exchange energy in the solid phase). At any temperature above the exchange energy scale, the system is well-represented by spinless electrons in the solid phase since the electrons stay far apart from each other.
    
   \section{Interpolation between the zero-temperature quantum ground state and the classical high-temperature thermodynamic state}\label{phonon}
   
   We have established, consistent with many earlier works, by exact numerical calculations that both finite-temperature classical and zero-temperature quantum 1D electron systems manifest a distinct effective 1D crystalline solid phase at high and low average electron densities, respectively.  Although there is no strict long-range order in a 1D system in the thermodynamic limit (destroyed by quantum and thermal fluctuations respectively in the quantum and the classical system), our results clearly demonstrate the existence of an effective finite-size Wigner crystal stabilized by Coulomb interaction.  Obviously, quantum and classical regimes must be smoothly connected in a physical system even though the effective crystal phase is preferred at low (high) average densities in the quantum (classical) case.  In the current section, we show how to establish the connection between quantum and classical regimes, and smoothly interpolate between them in order to obtain an effective temperature-density Wigner crystal crossover phase diagram for a 1D interacting electron system.
   
   We first introduce a simple model to connect the quantum ground state and the classical thermodynamic state. We note that for a 2D electron system, Hwang et al. theoretically obtained a quantum-classical crossover solid-liquid density-temperature phase diagram by appropriately matching the quantum Wigner crystal parameters to the corresponding classical limit \cite{plasmon2D}. The technique includes estimating the ratio of the average potential energy to the average kinetic energy (as a function of average density and temperature) and constraining the ratio to a predetermined constant in order to set the liquid-solid phase boundary. In this paper, we evaluate the solid-to-liquid phase crossover in a similar manner by  fixing the ratio between the vibration amplitude and the inter-electron distance, similar to the vibration of nucleus in a common lattice \cite{collectivemode}. This is not an absolute criterion of course (and indeed there cannot be any absolute criterion since strictly speaking there is no true 1D long-range order), but physically a vibration amplitude smaller (larger) than the inter-electron separation signifies a finite-size crystal (liquid). Our criterion is thus a generalization of the well-known Lindemann criterion for calculating the solid-liquid phase boundary of ordinary materials.
   
   We use the same binding potential with the general exponent $p$ as in Eq.\eqref{Hamiltonian}. However, the repulsive electron-electron interaction is taken to be purely Coulombic $1/|x_i-x_j|$ because we expand the oscillation around the classical equilibrium configuration of the Wigner crystal, where the electrons are essentially far apart from each other. The collective oscillation of the electrons is obtained through the phonon excitation spectrum. For this purpose, we first calculate the eigenmodes of the system, then address the thermal occupation of  each mode using the Bose-Einstein distribution. Thus, we are explicitly considering the “phonon spectra” of the effective 1D Wigner crystal by incorporating the external confinement (defining the finite system) and the inter-electron Coulomb interaction. When the phonon vibration amplitude (including the zero-point motion at $T=0$) is large, the crystal is considered to have `melted' into the liquid phase.
   
   We assume the set of particle positions to be $\{x_i\}$ and $x_1<x_2<...<x_N$ without any loss of generality. Because the binding potential is symmetric, we can set a constraint $x_1=-x_N$, and define the system size to be $L=x_N-x_1=2x_1$ as well as a new normalized coordinate $u_i=x_i/L$. In the new coordinates, $u_1=-u_N=0.5$ and $|u_i|<0.5$ $\forall~ 1<i<N$. Then the total potential energy is given by
   \begin{equation}
   	U=\frac{\hbar^2}{ma_B^2}\left[\frac{a_B}{L}\sum_{i<j} \frac{1}{u_i-u_j}+\frac{a_B^2}{L_0^2}\left(\frac{2L}{L_0}\right)^p\sum_i u_i^p\right].
   \end{equation}
   With a fixed set of $\{u_i\}$, $U$ is minimized as 
   \begin{equation}\label{eq9}
   	\frac{ma_B^2U}{(p+1)\hbar^2} \ge \left(\frac{2a_B}{L_0}\right)^{1+\gamma}\left(\sum_i u_i^p\right)^{\gamma}\left(\sum_{i<j} \frac{1}{u_j-u_i}\right)^{p\gamma},
   \end{equation}
   with $\gamma=1/(p+1)$. The RHS of the inequality \eqref{eq9} can be further minimized, giving a set of equilibrium positions $\{u_i^*\}$ independent of the size of the binding potential $L_0$. With this equilibrium set, the system size $L$ is given by
   \begin{equation}
   \begin{split}
   	&\frac{a_B}{pL}\sum_{i<j} \frac{1}{u_j^*-u_i^*}=\frac{a_B^2}{L_0^2}\left(\frac{2L}{L_0}\right)^p\sum_{i} u_i^{*p}\\
   &\Rightarrow L= a_B^{-\gamma}L_0^{1+\gamma}\left(\frac{1}{2^pp}\sum_{i<j} \frac{1}{u_j^*-u_i^*}\sum_{i} u_i^{*p}\right)^\gamma.
   \end{split}
   \end{equation}
   The squared eigenmode frequencies are obtained by diagonalizing the matrix $A$   defined by
   \begin{equation}\label{eq11}
   	\begin{split}
   	A_{m,n}&=\frac{\partial^2 U}{\partial x_m^* \partial x_n^*}\propto \frac{-2a_B}{L^3|u_m^*-u_n^*|^3}\propto L^{-3},\\
   	A_{m,m}&=\frac{\partial^2 U}{\partial x_m^{*2}}\propto \frac{a_B^2}{L_0^2}\left(\frac{2}{L_0}\right)^p(p-1)L^{p-2}pu_m^{*p-2}\\ &\quad+\frac{a_B}{L^3}\sum_{j\ne m}\frac{2}{|u_m^*-u_j^*|^3} \propto L^{-3}.
   	\end{split}
   \end{equation}
   As a consequence, we have 
   \begin{equation}
   \begin{split}
      	\omega_i =\omega_{0i} (\rho a_B)^{3/2},
   \end{split}
   \end{equation}
   where $\omega_{0i}$ is the $i^{th}$ mode frequency at $\rho a_B=1$ or $L=Na_B$. The average occupation in each mode is given by the Bose-Einstein distribution $n(\omega_i)=\left(\exp(\beta \omega_i)-1\right)^{-1}$. The average vibration amplitude of each electron around its equilibrium position is
   \begin{equation}\label{eq14}
   \begin{split}
       	q^2&=\frac{1}{N}\sum_i\frac{E_i}{m\omega_i^2}\\
       	 &=\frac{\hbar}{mN} \sum_i \frac{1}{(\rho a_B)^{3/2}\omega_{0i}}\left(\frac{1}{\exp(\beta \omega_i)-1}+\frac{1}{2}\right).
   \end{split}
   \end{equation}
   Adopting the Lindemann melting criterion, the system is assumed to melt when $q$  is comparable to the lattice period with the Lindemann factor $c$. Specifically, 
   \begin{equation}\label{eq15}
   	q^2=c^2(L/N)^2=c^2/\rho^2,
   \end{equation}
   where $c$ is a predetermined constant. The specific value of the dimensionless number $c$ is irrelevant for our purpose and in much of our following discussions (although the value of $c$ does determine the critical density/temperature for solid-to-liquid crossover for any given particle number). Obviously, for a crystal to be well-defined one expects $c\ll1$ since a vibration amplitude comparable to the lattice spacing implies a typical liquid rather than a solid.  The theory, however, cannot constrain the value of $c$, which would depend on the experimental details and may not be unique. In experiments, the choice of $c$ would determine the precise quantitative phase diagram. We emphasize, however, that the qualitative phase diagram in dimensionless parameters will be the same as what we obtain.
   \subsection{Classical limit}
   The system behaves classically when the spacing between the energy levels is much less than the thermal energy or $\beta \omega \ll 1$. Expanding Eq.~\eqref{eq14} in a Taylor series of $\beta \omega$ and plugging it into Eq.~\eqref{eq15}, we have
   \begin{equation}\label{eq16}
   \begin{split}
      	&q^2 = \frac{\hbar}{Nm}\sum_i \frac{k_BT}{(\rho a_B)^3\omega_{0i}^2}=\frac{c^2}{\rho^2}\\
      	&\Rightarrow \Gamma =\frac{\rho a_B}{T/T_B} = \frac{1}{N c^2}\left(\frac{\hbar}{ma_B^2}\right)^2\left(\sum_i \frac{1}{\omega_{0i}^2}\right),
   \end{split}
   \end{equation}
   where $T_B=\hbar^2/(ma_B^2k_B)$. The condition derived in Eq.~\eqref{eq16}, therefore, defines the classical liquid-solid phase boundary for the effective 1D Wigner crystal with the basic crossover line being a straight line in the density-temperature phase diagram. It is noted that in the classical limit $\Gamma=(\rho a_B)/(T/T_B)\approx \braket{V}/\braket{K}$, where $\braket{V}$ is the average Coulomb potential and $\braket{K}\sim T$ is the average kinetic energy. This classical limit should apply for $T\gg T_F$ where $T_F\sim  \rho^2$ is the Fermi temperature of the 1D system.
   \subsection{Quantum limit at zero temperature}
   As $T\to 0$, $\beta\omega \to \infty$, then
   \begin{equation}\label{eq17}
   \begin{split}
          &q^2=\frac{\hbar}{2Nm(\rho_c a_B)^{3/2}}\sum_i \frac{1}{\omega_{0i}}=\frac{c^2}{\rho_c^2}\\
          &\Rightarrow\rho_c a_B=4N^2 c^4\left(\frac{\hbar}{ma_B} \sum_i \frac{1}{\omega_{0i}} \right)^{-2}.
   \end{split}
   \end{equation}
   Here $\rho_c$ defines the critical average density for solid ($\rho<\rho_c$) and liquid ($\rho > \rho_c$) quantum Wigner crystallization condition at $T=0$. From Eq.~\eqref{eq17}, it is clear that the product $\rho q$ decreases with increasing $\rho$. Thus, the vibration amplitude of the quantum crystal becomes larger compared with the lattice spacing as the average density increases leading to the preferential melting at higher average densities. This is consistent with our earlier conclusion in the quantum case. 
     \begin{figure}
     	\begin{minipage}{0.2cm}
     		\rotatebox{90}{$T/T_B$}
     	\end{minipage}
     	\begin{minipage}{0.42\textwidth}
     		\centering
     		\includegraphics[width=0.9\textwidth]{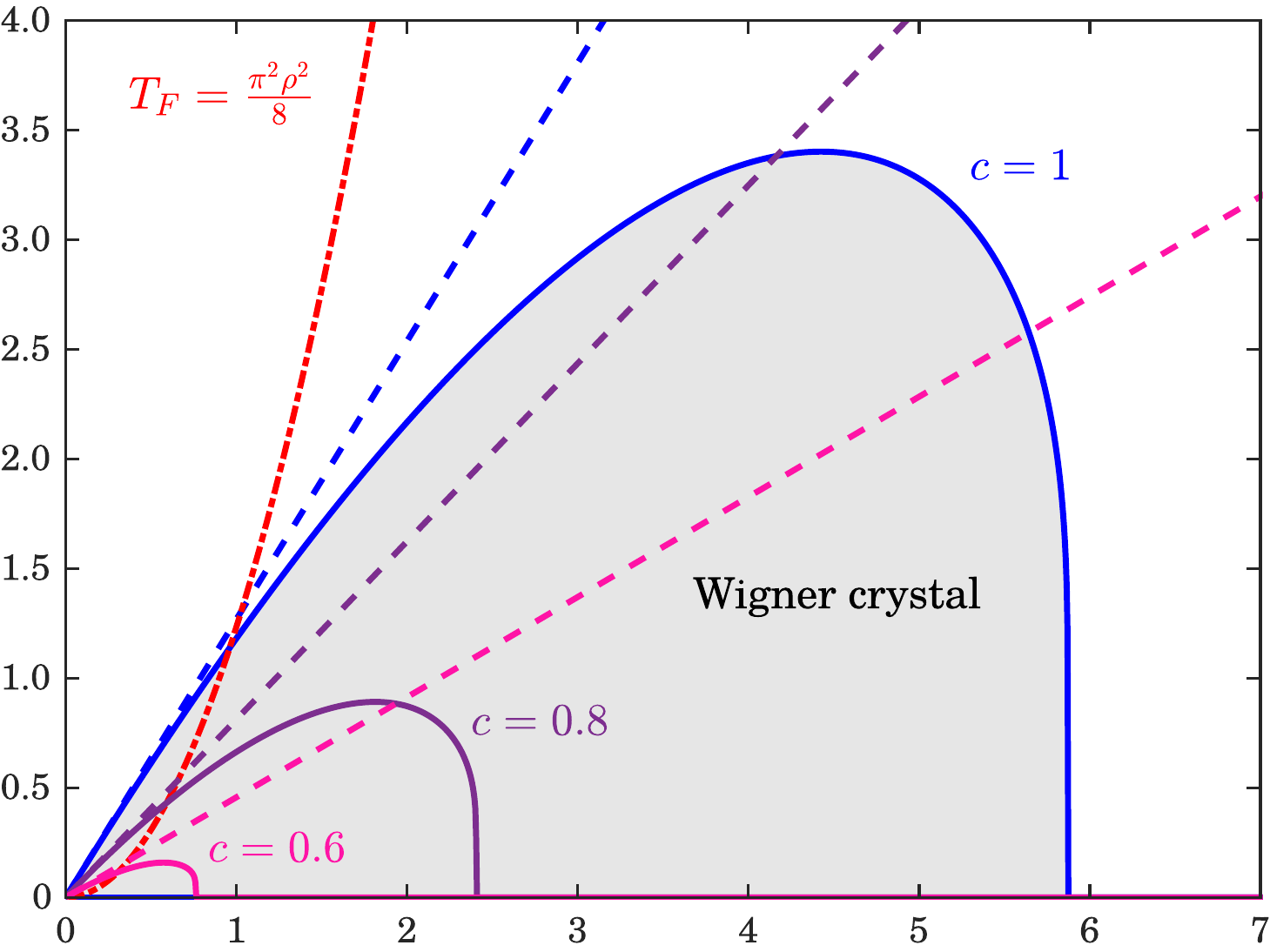}\\
     		 $\rho a_B$
     	\end{minipage}
     	\caption{Phase diagram of 1D system for $N=4$ electrons. The dashed and solid lines of the same color represent the classical limit given by Eq.~\eqref{eq16} and the semi-classical Wigner crystal boundary given by Eqs.~\eqref{eq14}-\eqref{eq15} and  at a particular choice of the parameter $c$. The red dash-dotted line is the Fermi temperature. When the temperature is much larger than the Fermi temperature, the phase transition lines approach the corresponding classical limits as they should.}	\label{fig8}
     \end{figure}
   
   In Fig.~\ref{fig8}, we show our calculated effective phase diagram in a small system of $N=4$  electrons by directly numerically solving Eq.~\eqref{eq14} as a function of average density (defined by the binding potential) and temperature. We present the calculation with the choices of $c=0.6,0.8$ and 1 for this figure - different values of $c$ give precisely the same qualitative phase diagram. The shaded region has $q<c/\rho$ and can be considered  the effective 1D Wigner crystal. At high average densities, the system melts due to quantum fluctuations whereas at high temperatures the system melts due to thermal fluctuations. We also plot the classical limit as calculated from Eq.~\eqref{eq16} as dashed lines. The full solution approaches the classical limit when the temperature is larger than the Fermi temperature given by $T_F=\pi^2 \rho^2/8$. At very low average densities, as expected, classical and quantum solutions agree. Note that in Fig.~\ref{fig8}, the pure classical Wigner crystal regime is rather small (the low-density regime between the blue and red lines). The fragility of a classical 1D Wigner crystal phase arises from the fact that the existence of a classical crystal requires very low average density corresponding to very low Fermi temperature - thus the classical crystal is constrained by the Fermi temperature on the one hand (indicating classical to quantum crossover) and the low melting temperature on the other hand (indicating the solid to liquid crossover). This fragility of the classical Wigner crystal was also found to be the case in a completely different calculation \cite{phase_montecarlo} employing the static structure factor as the diagnostic to distinguish between the classical and the quantum regime in contrast to our use of the Lindemann criterion. In contrast to Ref. \cite{phase_montecarlo}, however, which finds two disjointed classical solid phases, we find only a small sliver of a classical effective Wigner phase between the quantum Wigner crystal and the classical liquid.
   
   It should be noted that through Eq.~\eqref{eq11}, $\omega_{0i}$ depends on the exponent of the binding potential through the term $(p-1)pu^{* p-2}$. For $|u^*|<0.5$, this term decreases with higher exponent $p$, leading to lower $\omega_{0i}$; thus smaller $\rho_c$ and larger $\Gamma$. The exponent $p$ controls the steepness of the external binding potential (which defines the 1D confinement), i.e. lower $p$ means steeper potential.  Intuitively, when the potential is steeper, the particles are drawn more strongly towards the center, and as a result, closer to each other, making the Wigner crystal harder to form since the effective average density in the bulk of the 1D system becomes larger even for the same nominal system size and electron number.  The details of the quantum  confinement defining the 1D system thus also play a direct role in the effective Wigner crystallization phenomenon.
  
   \subsection{Long-range order for larger systems}
\begin{figure}
	\begin{minipage}{0.2cm}
		\rotatebox{90}{\hspace{0.2in} Dimensionless $\Gamma$ and $ a_B\rho_c$}
	\end{minipage}
	\begin{minipage}{0.42\textwidth}
		\centering
		\includegraphics[width=\textwidth]{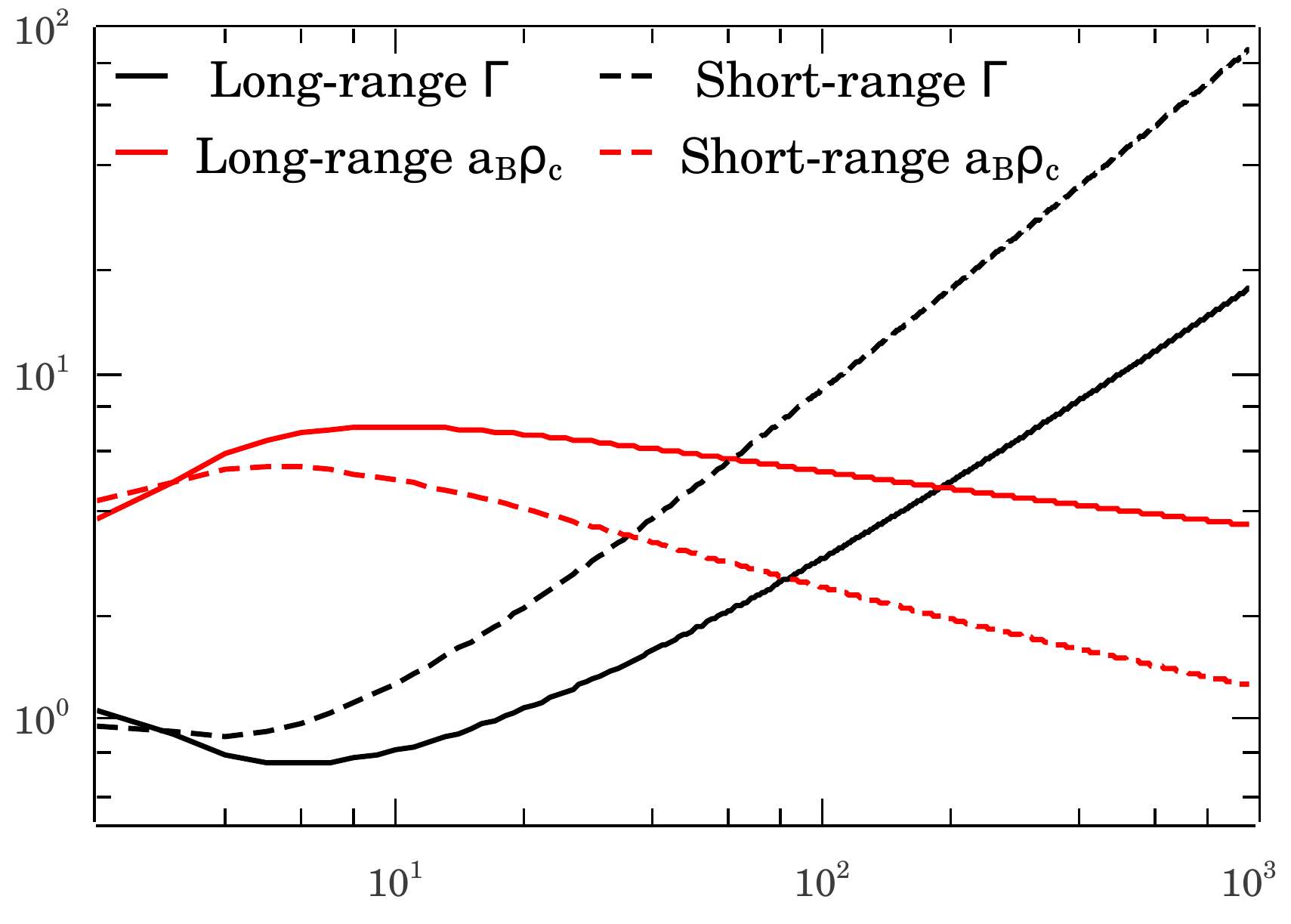}
		\hspace*{0.8cm} $N$
	\end{minipage}
	\hspace*{0.5cm}
	\caption{Dimensionless $\Gamma$ and $ a_B\rho_c$ as functions of the electron number $N$ for long-range and short-range interacting systems. The Wigner phase shrinks when $\Gamma$ increases and $\rho_c a_B$ decreases. We choose $c=1$ in this figure.} 
		\label{fig15}
\end{figure}

   So far, we have discussed the formation of an effective 1D Wigner crystal in a finite system by using various criteria for the spatial charge density distribution and correlation as well as the vibration amplitude for the localized phonon eigenmodes of the finite system.  The question now arises on how the physics of the finite system effective Wigner crystallization is modified as the system size increases. We previously argue that for a finite system, both long-range and short-range (e.g. screened Coulomb as through gating) interactions can induce a Wigner-crystal order. On the other hand, we know that in the $N\to \infty$ limit, the Wigner phase obviously disappears regardless of the interaction type in 1D systems.  The order is destroyed by quantum fluctuations for $T=0$ and by thermal fluctuations for nonzero $T$.  However, the form of interaction may be critical to the rate of Wigner order disappearance as the system becomes infinite. To establish this point, we plot the Wigner crystal phase boundary characterized by $\Gamma$ and $\rho_c$ in Fig.~\ref{fig15} at increasing number of electrons $N$ for two types of interactions (i) the long-range Coulomb interaction and (ii) the short-range nearest interaction $V_{sr}(x_i,x_j)=1/|x_j-x_i|$ for $|j-i|=1$ and $V_{sr}=0$ otherwise, with $i,j$ are the spatial order indices.
   
   For small systems, the Wigner crystal phase changes non-trivially with an increasing number of electrons. The crystal phase first expands (i.e. decreasing $\Gamma$ and increasing $\rho_c$) from $N=2$ to $N=6$, then shrinks at larger values of $N$. Our numerical simulations indicate that the optimal size $N^*$ where the crystal region is maximum decreases as the exponent $p$ of the binding potential increases. Specifically, $N^*=8$ for $p=2$ and $N^*=2$ for $p\ge 8$. These details do not depend on the choice of $c$, but $\rho_c$ and $T_c$ (for a specific $\rho<\rho_c$) do depend on the choice of $c$.
   
   For large $N$, the Wigner phase reduces monotonically with $N$ (decreasing $\rho_c$ and increasing $\Gamma$). To study the behavior in the limit $N\to \infty$, we assume that the specific form of the trapping potential only affects the first few modes and the eigenvector of the $n^{th}$ mode of Eq.~\ref{eq11} can have the form $u_i=\sin[\pi n(i-1) /(N-1)]$, similar to an infinite square well potential. Substituting this trial solution into Eq.~\ref{eq11}, we can estimate
   \begin{equation}
   \begin{split}
     \omega_{0n}^2 &= \left(\frac{2\hbar}{ma_B^2}\right)^2\sum_{i=1}^{\infty} \frac{1-\cos(i\pi n/(N-1))}{i^3}\\
     &\approx 2\left(\frac{\hbar}{ma_B^2}\right)^2 \left(\frac{\pi n}{N}\right)^2\ln\left(\frac{N}{\pi n}\right).
   \end{split}
   \end{equation}
   We can approximate the effect of the system  size by evaluating $\Gamma$ and $\rho_c a_B$ (in Eq.~\ref{eq16}) in the large$-N$ limit 
   \begin{equation}\label{longrange}
   \begin{split}
     &\Gamma =\frac{1}{Nc^2}\left(\frac{\hbar}{ma_B^2}\right)^2\sum \frac{1}{\omega_{0i}^2} \sim \frac{N}{\ln N};\\
     &\rho_c a_B = 4N^2 c^4\left(\frac{\hbar}{ma_B} \sum_i \frac{1}{\omega_{0i}} \right)^{-2} \sim \frac{1}{\ln N}.
   \end{split}
   \end{equation}
   
   Proceeding similarly to the long-range case, we find for the nearest interacting case that $\omega_{0n}^2 \propto (\pi n/N)^2$ and thus
   \begin{equation}\label{shortrange}
   	\begin{split}
   	 \Gamma_{sr} \sim N;\quad \rho_{c,sr}a_B \sim \frac{1}{\ln^2 N}.
   	\end{split}
   \end{equation}
   Equations.~\eqref{longrange} and \eqref{shortrange} together with Fig.~\ref{fig15}, demonstrate the fact that there can be no quantum Wigner crystal as the system size and the number of particles both go to infinity (keeping the density constant). Thus, an effective 1D Wigner crystal is readily observable in small systems, but does not exist in very large systems! However, the disappearance of the long-range order in the thermodynamic limit of a Coulombic system is much slower than a system with short-range interaction as predicted by the Luttinger liquid theory \cite{Schulz}. In addition, since $\Gamma$ and $\rho_c$ of the long-range Coulomb system vary with $\ln N$, effective long-range spatial order persists to rather large system sizes in the 1D Coulomb interacting system ($\rho_c a_B \approx 6.0$ for $N\sim 50$) enabling the clear observation of an effective 1D Wigner crystal crossover phase up to rather large system sizes even at finite temperatures. This remarkable effect is due to the slowly decaying law $e^{-C\sqrt{\ln x}}$ of $4k_F$ oscillation in a Coulomb Luttinger liquid. This slow decay of crystalline order for the Coulomb system makes the 1D Wigner crystal phase apparently similar to 2D and 3D Wigner crystals, but the fact remains that in 1D we only have an effective Wigner crystal existing only in finite systems, a distinction not often emphasized in the literature.
   
   \section{Conclusion}\label{conclude}
    We have theoretically studied by exact numerical techniques for small 1D Coulomb interacting systems the spatial density distribution, the density correlation, and the eigenmode vibration properties at zero and finite temperatures in both quantum and classical limits. For the zero-temperature case, we show that the effective quantum crystal characteristics associated with distinct spatial density peaks are strongly enhanced when the average electron density decreases. Furthermore, the crystal-like density correlations associated with the slowly decaying $4k_F$ oscillations are present at all average densities with the oscillation amplitude decreasing with increasing average density. For high temperatures, by using the exact classical partition function, we show that the spatial density profile exhibits the same $N$-crest signature as in the quantum case except now the crest amplitudes decrease with increasing temperature or decreasing average density, in contrast to the zero-temperature quantum limit. We also show that in both cases the density correlation is a better indication of the Wigner crystal than the spatial density modulation.
   
   One important result of this paper is that we can connect the two limits for the effective 1D Wigner crystallization (and interpolate smoothly between the quantum and the classical regime) using a model of excited phonons to calculate the vibration amplitudes of the localized electron motion, and consequently using the Lindemann melting criterion to define the solid-liquid transition. We obtain the resultant 1D effective finite-size Wigner crystal phase diagram in the density-temperature space, finding that there exists an isolated density-temperature region where the system exists in an effective 1D Wigner crystal phase for small system sizes. We find that the quantitative aspects of the effective Wigner crystallization depend crucially on the details of the confinement potential creating the 1D system as well as on the precise Lindemann criterion (i.e. what fraction of the lattice constant the phonon amplitude can equal before melting into the liquid phase), but the qualitative phase diagram is universal.
   
   We also show how this crystal region in the phase diagram shrinks very slowly as the number of particles increases. Notably, even though as we show, short-range interactions can induce the effective Wigner crystallization, this Wigner phase disappears much faster than that induced by the true long-range Coulomb interaction with increasing system size. This result may help guide future experiments in searching for 1D Wigner crystal in 1D systems with a finite number of electrons.  Our obtained effective phase diagram should also be directly relevant to experiments. In particular, our predicted thermal melting of the 1D Wigner crystal should be directly observable in the experimental set up of Ref.~\cite{exp0} where raising the temperature at a fixed average density (i.e. fixed electron numbers in the nanotube of a fixed length) should lead to a strong suppression of the spatial density peaks in accordance with our qualitative phase diagram with lower average density manifesting a lower melting temperature. We emphasize that although the quantitative details (e.g. the precise melting temperature) of the effective 1D Wigner crystal depend on many parameters not amenable to theoretical analyses (e.g. the value of $c$), the qualitative form of our calculated classical-quantum phase diagram should remain valid.
   
   We conclude by emphasizing the important theoretical results obtained in our work: (1) the effective finite-size 1D Wigner density modulation exists at all densities for spinless (or polarized) systems, this structure, however, is suppressed by the degeneracy at low on-site interaction;  (2) the long-distance decay of the crystalline order is extremely slow, enabling a clear experimental observation of the effective 1D solid phase up to many electrons; (3) the liquid to solid crossover is characterized by a strong suppression of the exchange energy, which vanishes in the solid phase for all practical purposes making the 1D Wigner crystal spin incoherent; (4) as $N$ increases, the classical melting temperature and the ``critical'' density of 1D Coulombic Wigner crystal reduce much slower than a short-range interacting system as constrained by Coulomb Luttinger liquid theory, this qualitative finding should remain valid independent of our approximations; (5) the actual definition of the effective 1D Wigner crystal depends on the experimental resolution determining the density variations, but in general, extending the interaction range by isolating from external gates should enhance the density variations in the system even at a fixed $r_s$ (or average density) as shown in Fig.~\ref{fig_shortrange}. Although some of these conclusions (in particular items 1-3) are known, our results put precise quantitative perspectives on these qualitative conclusions making the somewhat vague concept of 1D Wigner crystallization on a more concrete footing.   
   
   In this paper, we consider a strictly 1D system with only one active 1D channel in the system. In reality, if this condition is relaxed, there is a possibility of chain modes excitation associated with transverse motion \cite{R7,R8}. We obtain in Appendix A a more detailed condition for which the transverse modes are suppressed and the system can be considered as strict 1D. In case most of this condition is violated, a more careful treatment is needed but we believe that our conclusions still hold qualitatively.
   
   \section*{acknowledgments}
   	We thank F. Peeters for bringing Refs. \cite{R2} and \cite{R8} and G. Zar\'{a}nd for bringing Ref. \cite{R6} to our attention. We thank Dr. Domenico Giuliano
   	for a number of useful suggestions. This work is supported by the Laboratory for Physical Sciences.
   
 \appendix
 \begin{widetext}
 	 \section{Transverse mode excitation}
    In this appendix, we justify our use of the $1/x$ interaction despite considering a 1D system. Moreover, we show that in the context of our paper, it is not necessary to include excited transverse mode. The field can be decomposed as $\Psi(\vec{r}) = \psi_l(x)\psi(x,\vec{\rho})$ where $\psi_l$ is the longitudinal field, $\psi$ is the transverse field, $x$ is the longitudinal coordinate and $\vec{\rho}$ is the transverse coordinate including the the radius $\rho$ and the angle $\theta$. To obtain an analytic expression, we use a quadratic transverse background potential
 	\begin{equation}
 	V(\vec{\rho}) = \frac{\hbar^2}{2md^2}\left(\frac{\rho}{d}\right)^2,
 	\end{equation}
 	where $d$ is roughly the transverse size of the tube. Then $\psi(\vec{\rho})$ can be expanded by eigenstates of the transverse potential $\psi(x,\vec{\rho}) = \sum_{m,n} \psi_{n,m}(\vec{\rho}) c(x)_{n,m}$. We have the usual commutation relation $\{ c_{\alpha}^\dagger(x),c_{\beta}(x')\} = \delta_{\alpha,\beta}\delta(x-x')$ where we denote $\alpha=(n,m)$ for conciseness. We show here two lowest transverse eigenstates
 	\begin{equation}
 	\begin{split}
 	\psi_{(0,0)}(\vec{\rho}) = \frac{1}{\sqrt{\pi}d}e^{-\frac{\rho^2}{2d^2}},\quad \psi_{(1,\pm1)}(\vec{\rho}) = \frac{1}{\sqrt{\pi}d}\frac{\rho}{d} e^{-\frac{\rho^2}{2d^2}}e^{\pm i\theta}.
 	\end{split}
 	\end{equation}
 	The Coulomb density-density interaction is given by
 	\begin{equation}
 	\begin{split}
 	H_{int}=\frac{\hbar^2}{2ma_B}\int d\vec{r}_1 d\vec{r}_2 \frac{n_l(x_1)n_l(x_2)}{\sqrt{(x_1-x_2)^2+(\vec{\rho}_1-\vec{\rho}_2)^2}} &\sum_{\alpha_1,\alpha_2,\alpha_3,\alpha_4} \psi^*_{\alpha_2}(\vec{\rho_2})\psi^*_{\alpha_1}(\vec{\rho_1})\psi_{\alpha_3}(\vec{\rho_1})\psi_{\alpha_4}(\vec{\rho_2})\\
 	& \quad \times c^\dagger_{\alpha_2}(x_2)c^\dagger_{\alpha_1}(x_1)c_{\alpha_3}(x_1)c_{\alpha_4}(x_2),
 	\end{split}
 	\end{equation}
 	where $n_l$ is the longitudinal density. We now expand the interaction in terms of transverse mode indices. To evaluate the integral, we use the Fourier representation of the Coulomb potential
 	\begin{equation}
 	\frac{1}{|\vec{r}|} = 2\int dk e^{ikx}\int\frac{dq^2}{(2\pi)^2}\frac{e^{i\vec{q}\cdot\vec{\rho}}}{k^2+q^2} 
 	\end{equation}
 	For $\alpha_1=\alpha_2=\alpha_3=\alpha_4=0$ and $z=x_1-x_2$,
 	\begin{equation}
 	\begin{split}
 	H_{int,0} &=\frac{\hbar^2}{2ma_B}\int dx_1dx_2n_l(x_1)n_l(x_2)\int dk2e^{ikz}\int \frac{dq^2}{(2\pi)^2} \frac{e^{-q^2d^2/2}}{q^2+k^2}\\
 	&=\frac{\hbar^2}{2ma_B}\int dx_1dx_2n_l(x_1)n_l(x_2)\int dke^{ikz}\frac{-e^{k^2d^2/2}\text{Ei}(-k^2d^2/2)}{2\pi}\\
 	&=\frac{\hbar^2}{2ma_B}\int dx_1dx_2n_l(x_1)n_l(x_2) \sqrt{\frac{\pi}{2d^2}} e^{\frac{z^2}{2d^2}}\text{erfc}\left[\frac{|z|}{\sqrt{2}d}\right],
 	\end{split}
 	\end{equation}
 	where we have omitted the creation/annihilation operators for brevity. Note that if we assume no excitation in the transverse mode $\braket{c^\dagger_{0}(x_2)c^\dagger_{0}(x_1)c_{0}(x_1)c_{0}(x_2)} = 1$ and the term
 	\begin{equation}
 	\sqrt{\frac{\pi}{2d^2}} e^{\frac{z^2}{2d^2}}\text{erfc}\left[\frac{|z|}{\sqrt{2}d}\right] = \begin{cases}
 	\sqrt{\frac{\pi}{2}}\frac{1}{d} \text{ for } z\to 0\\
 	1/|z| \text{ for } z\to \infty
 	\end{cases}.
 	\end{equation}
 	So the interaction term $1/\sqrt{(x_1-x_2)^2+d^2}$ is a reasonable approximation if we assume no transverse mode excitation. The lowest-order term that can excite the transverse mode $\propto c^\dagger_{(1,1)}(x_2)c^\dagger_{(1,-1)}(x_1)c_{0}(x_1)c_{0}(x_2)$ (the term $\propto c^\dagger_{(1,1)}(x_2)c^\dagger_{0}(x_1)c_{0}(x_1)c_{0}(x_2)$ vanishes due to rotational symmetry). The amplitude of this term is 
 	\begin{equation}
 	\begin{split}
 	H_{int,1} &= \frac{\hbar^2}{2ma_B}\int dx_1dx_2n_l(x_1)n_l(x_2)\int dk2e^{ikz}\int \frac{dq^2}{(2\pi)^2} \frac{q^2d^2e^{-q^2d^2/4}}{4(q^2+k^2)}\\
 	&= \frac{\hbar^2}{2ma_Bd}\int dx_1dx_2 n_l(x_1)n_l(x_2)\left\{\frac{-2|z|}{d} + e^{z^2/d^2}\sqrt{\pi}\left[1+\frac{2z^2}{d^2}\text{erfc}\left(\frac{|z|}{d}\right)\right]\right\}.
 	\end{split}
 	\end{equation}
 	where $z=x_1-x_2$. Suppose the density distribution has the form of equally spaced peaks with distance $a$. Then the transverse excitation term strength is
 	\begin{equation}
 	H_{int,1} \approx \frac{\hbar^2}{2ma_Bd}\left\{\frac{-2a}{d} + e^{a^2/d^2}\sqrt{\pi}\left[1+\frac{2a^2}{d^2}\text{erfc}\left(\frac{a}{d}\right)\right]\right\}.
 	\end{equation}
 	The transverse mode gap is given by $\Delta E = \hbar\omega=\hbar^2/(md^2)$. Then  the ratio between the coupling strength and the gap is
 	\begin{equation}
 	\frac{H_{int,1}}{\Delta E} = \frac{d}{2a_B}\left\{\frac{-2a}{d} + e^{a^2/d^2}\sqrt{\pi}\left[1+\frac{2a^2}{d^2}\text{erfc}\left(\frac{a}{d}\right)\right]\right\}.
 	\end{equation}
 	This ratio vanishes when either $d\ll a_0$ ($r_s\eta \ll 1/N$) or $d\ll a$ ($\eta\ll 1/N$).
 	\normalcolor
 	\section{Additional simulation results}
 	
 	 We present the simulations for smaller spinful systems, namely systems of 4 and 2 particles. The observations discussed earlier apply also in these smaller systems. For $N=2$ and 4, by decreasing $r_s$ or increasing $\eta$, one can induce crossover from the Wigner crystal phase to the liquid phase. Phenomenologically, this phase transition is marked by the spreading of spatial density peaks for spinless electrons and the merging of the density peaks for spinful electrons. 
 	
  \begin{figure*}[ht!]
 	\centering
 	\begin{minipage}{0.2cm}
 		\rotatebox{90}{\hspace*{0.5cm} $\rho(x)L_0$ }
 	\end{minipage}
 	\begin{minipage}{0.95\textwidth}
 		\centering
 		\includegraphics[width=0.32\textwidth]{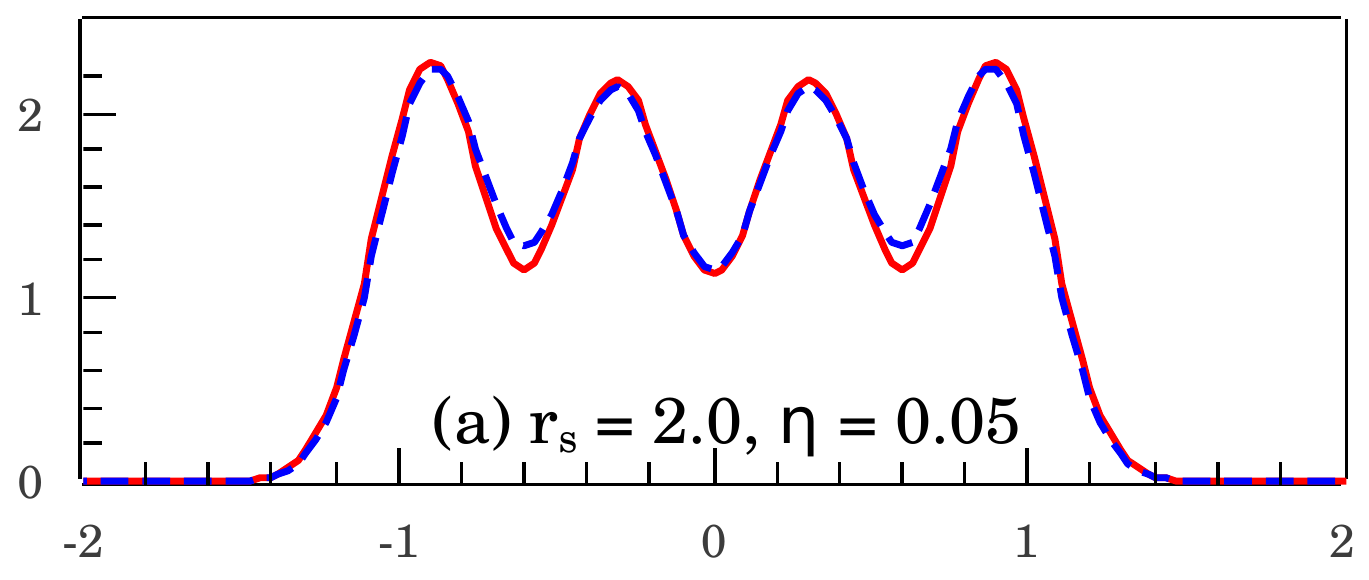}
 		\includegraphics[width=0.32\textwidth]{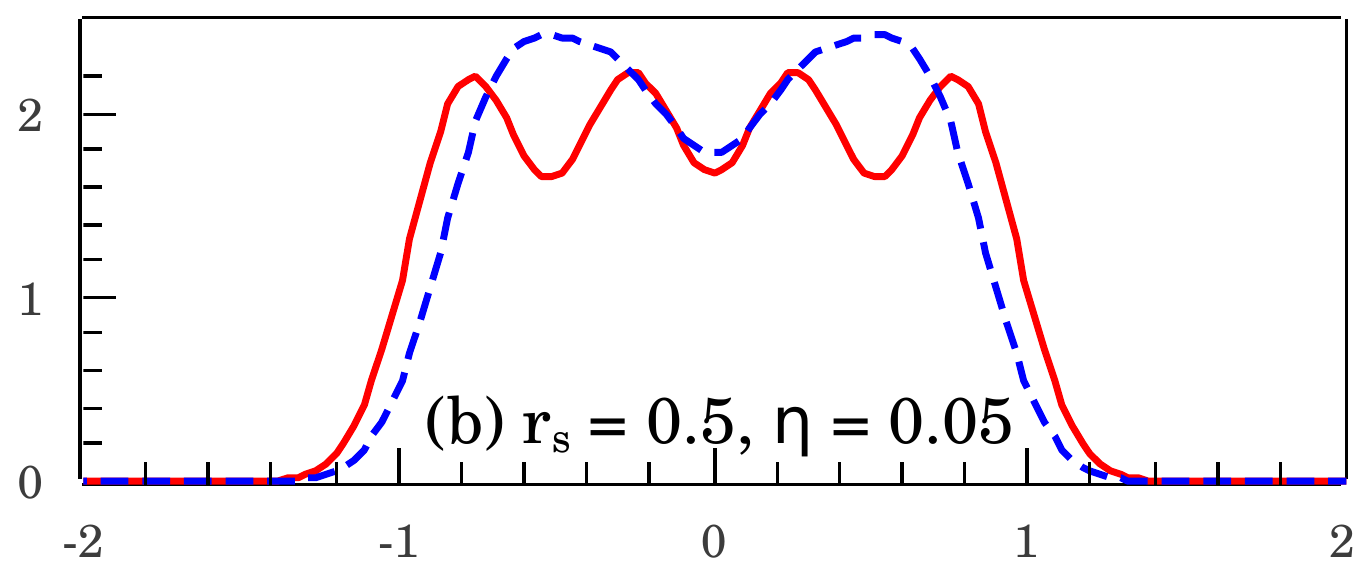}
 		\includegraphics[width=0.32\textwidth]{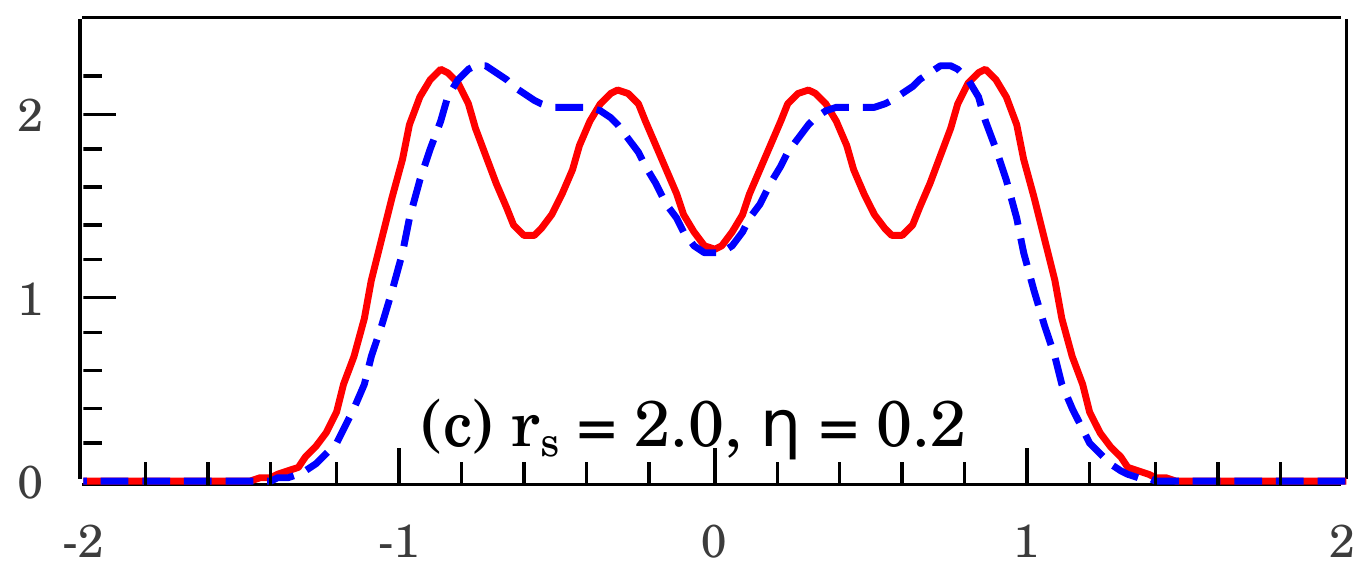} 	
 	\end{minipage}
 	$x/L_0$\\
 	\vspace{0.1in}
 	\begin{minipage}{0.2cm}
 		\rotatebox{90}{\hspace*{0.2cm}FFT$[\rho(x)L_0]$}
 	\end{minipage}
 	\begin{minipage}{0.95\textwidth}
 		\centering
 		\includegraphics[width=0.32\textwidth]{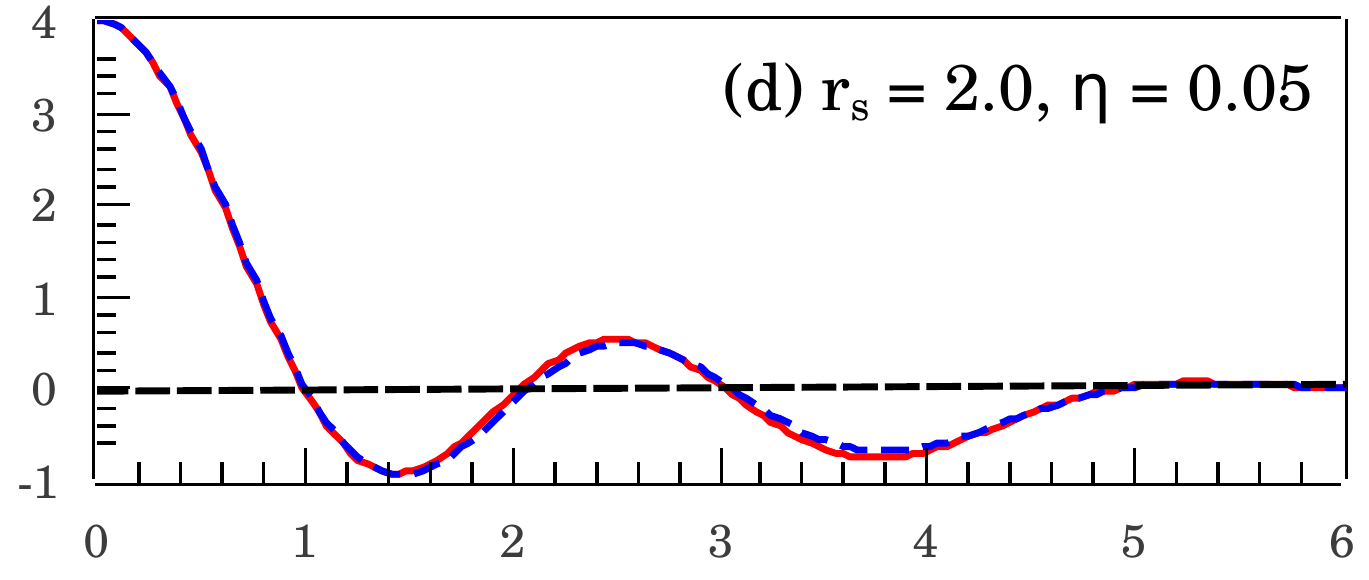}
 		\includegraphics[width=0.32\textwidth]{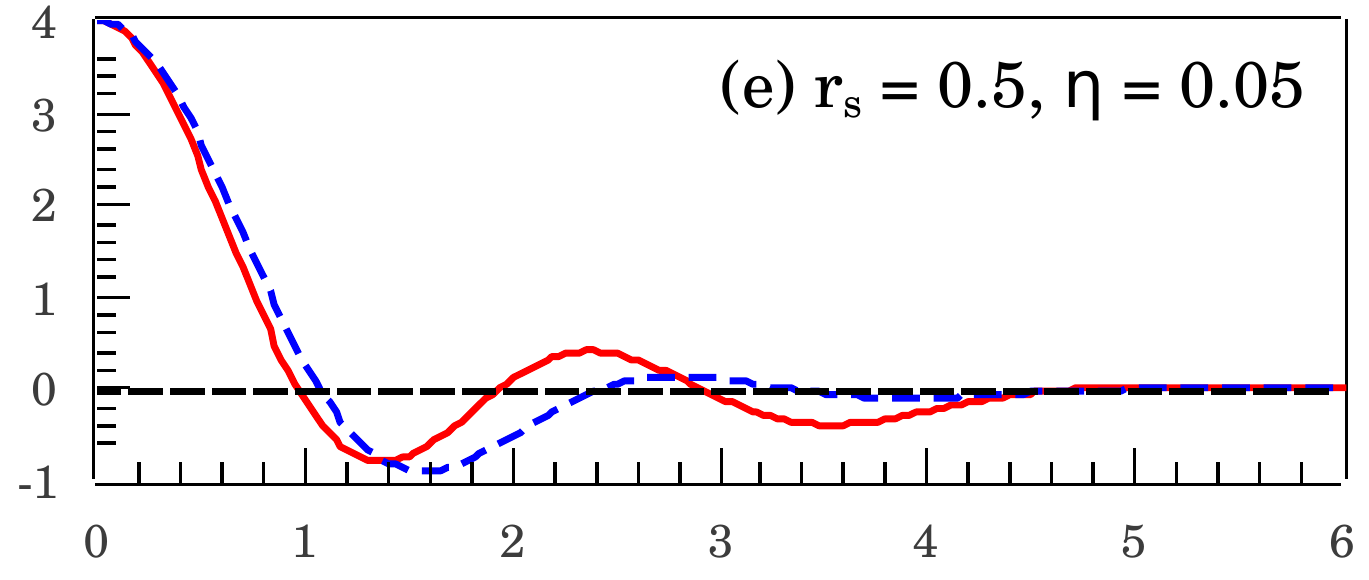}
 		\includegraphics[width=0.32\textwidth]{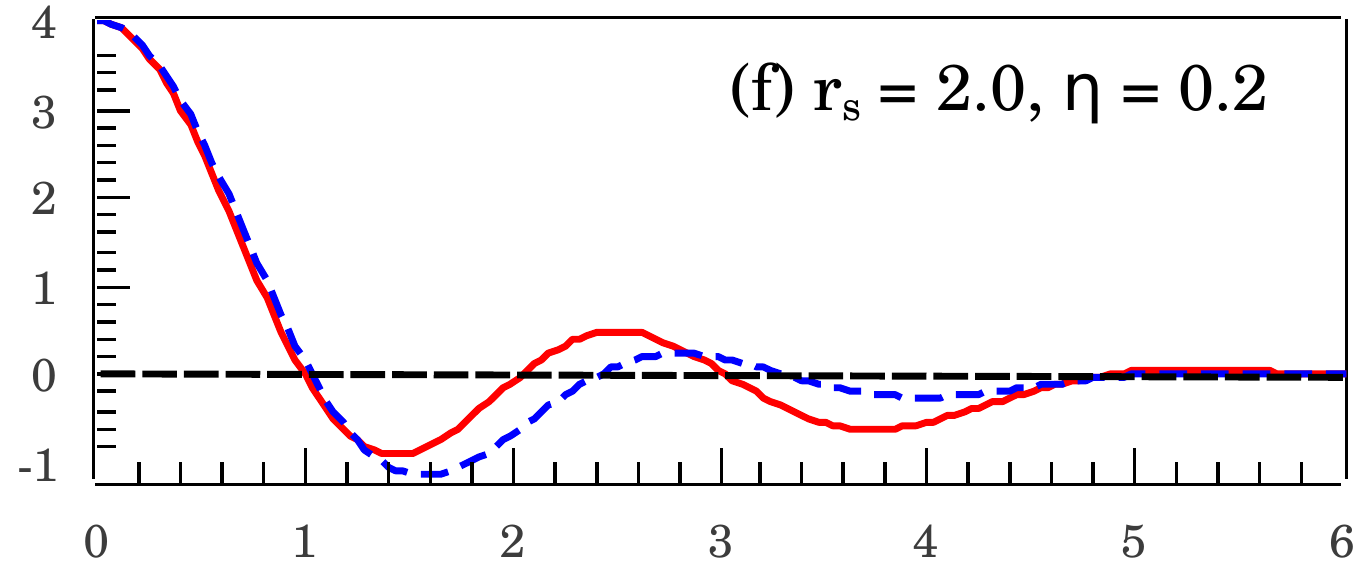}	
 	\end{minipage}
 	$k/k_F$
 	\caption{Simulation results for 4-particle spinful (blue dash) and spinless (red line) systems. The right column shows the Wigner crystal phase occurring at a high $r_s$ and a low $\eta$, the middle column shows the liquid phase induced by decreasing $r_s$, the right column shows the liquid phase at a higher $\eta$.}\label{fig10a}
  \end{figure*} 		
 	   \begin{figure*}[ht!]
 		\centering
 		\begin{minipage}{0.2cm}
 			\rotatebox{90}{\hspace*{0.4cm} $\rho(x)L_0$}
 		\end{minipage}
 		\begin{minipage}{0.95\textwidth}
 			\centering
 			\includegraphics[width=0.32\textwidth]{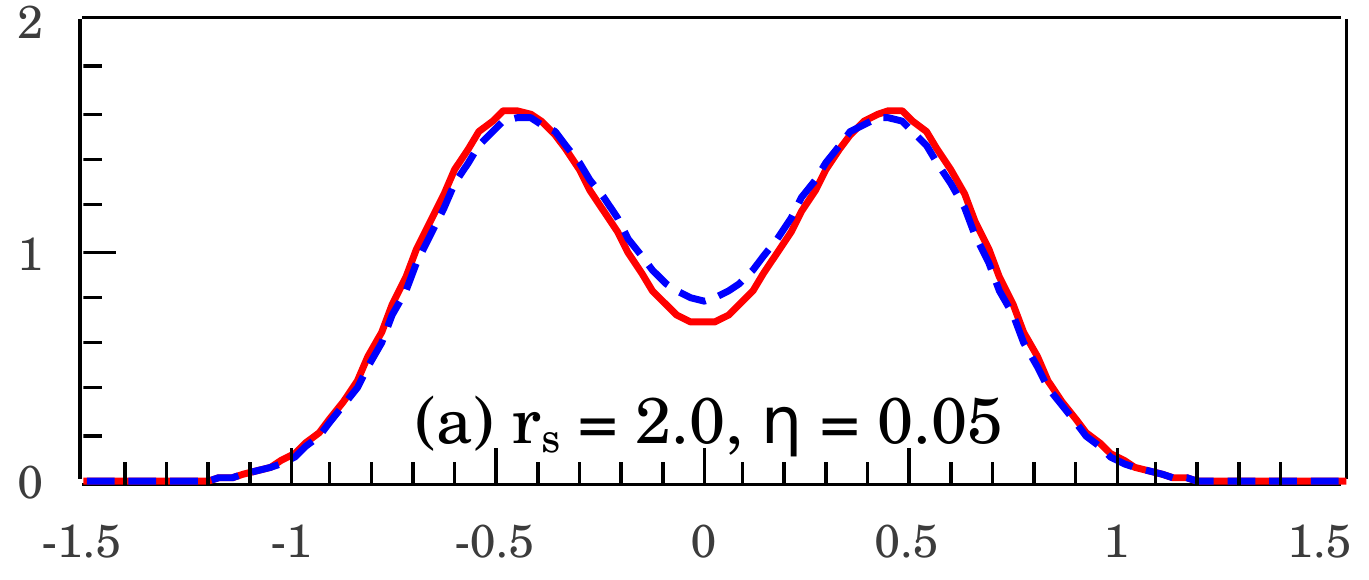}
 			\includegraphics[width=0.32\textwidth]{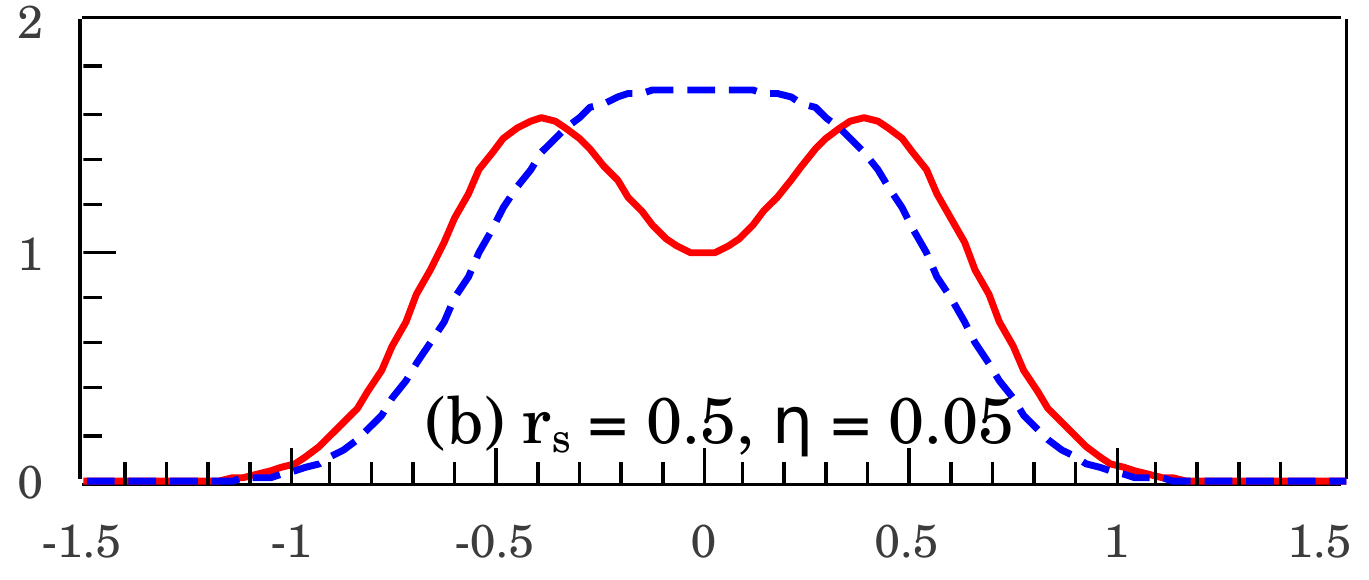}
 			\includegraphics[width=0.32\textwidth]{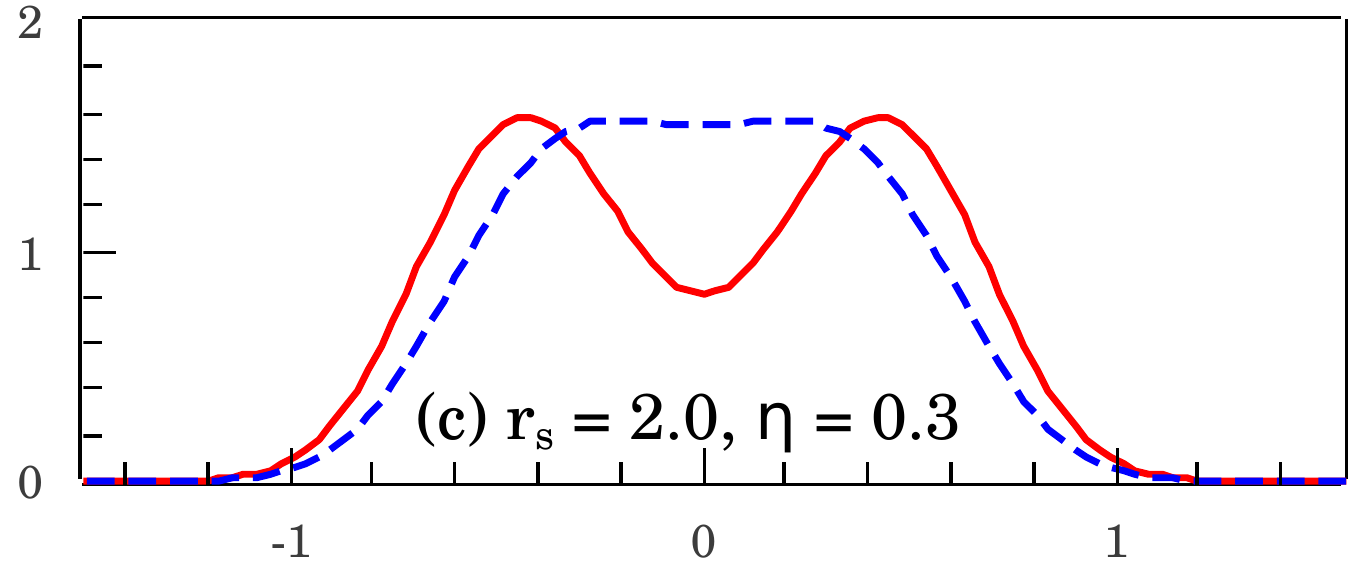} 	
 		\end{minipage}\\
 		$x/L_0$
 		\caption{Simulation results for 2-particle spinful (blue dash) and spinless (red line) systems in (a) Wigner crystal phase, (b) liquid phase induced by decreasing $r_s$ and (c) liquid phase induced by increasing $\eta$. The Fourier transform is not presented due to the lack of periodicity.}\label{fig11a}
 	\end{figure*}

 \end{widetext}
      
   \bibliographystyle{apsrev4-1}
   \bibliography{refs}
   
\end{document}